\documentclass[12pt]{article}
\pdfoutput=1

\newlength{\abstractwidth}
\usepackage{cite}
\usepackage{mathtools}
\usepackage{dsfont}
\usepackage{amssymb}
\usepackage{latexsym}
\usepackage{braket}
\usepackage{graphicx}
\usepackage{subcaption}
\usepackage[utf8]{inputenc}
\usepackage[usenames]{color}

\usepackage{epsf}
\usepackage{color}
\usepackage{graphicx}
\usepackage{tikz}
\usepackage{dsfont}

\usetikzlibrary{arrows,shapes,positioning}
\usetikzlibrary{decorations.markings}
\usepackage[rightcaption]{sidecap}
\tikzstyle arrowstyle=[scale=1]
\tikzstyle directed=[postaction={decorate,decoration={markings,
    mark=at position .65 with {\arrow[arrowstyle]{stealth}}}}]
\tikzstyle reverse directed=[postaction={decorate,decoration={markings,
    mark=at position .65 with {\arrowreversed[arrowstyle]{stealth};}}}]
\usetikzlibrary{positioning}

\usepackage{hyperref}
\definecolor{darkred}{rgb}{0.8,0.1,0.1}
\hypersetup{colorlinks=true, linkcolor=darkred, citecolor=blue, linktoc=page, bookmarksnumbered}

\usepackage[shadow,textwidth=2cm]{todonotes}
\usepackage{ifthen}
\reversemarginpar
\newcounter{todocounter}

\colorlet{jgcolor}{green!40!white}

\newcommand{\jginline}[2][]{
  \ifthenelse { \equal {#1} {} }
    { \def\temp {#2} }  
    { \def\temp {#1} }   
  \stepcounter{todocounter}\todo[color=jgcolor,inline,caption={\textbf{\thetodocounter. JG} \temp}]{\textbf{\thetodocounter. JG:} #2}{}}

\colorlet{jkcolor}{blue!20!white}

\newcommand{\jkinline}[2][]{
  \ifthenelse { \equal {#1} {} }
    { \def\temp {#2} }  
    { \def\temp {#1} }   
  \stepcounter{todocounter}\todo[color=jkcolor,inline,caption={\textbf{\thetodocounter. JK} \temp}]{\textbf{\thetodocounter. JK:} #2}{}}

\newcommand\narrowdots{\hspace{-0.1em}...}

\renewcommand{\thefootnote}{\fnsymbol{footnote}}
\renewcommand{\thanks}[1]{\footnote{#1}}
\newcommand{\starttext}{
\setcounter{footnote}{0}
\renewcommand{\thefootnote}{\arabic{footnote}}}
\numberwithin{equation}{section}
\newcommand{\bea}{\begin{eqnarray}}
\newcommand{\eea}{\end{eqnarray}}

\newcommand{\bma}{\begin{matrix}}
\newcommand{\ema}{\cr\end{matrix}}

\def\cA{{\cal A}}
\def\cB{{\cal B}}
\def\cC{\,{\cal C}\!}

\def\cF{{\cal F}}
\def\cG{{\cal G}}

\def\ZZ{{\mathbb Z}}
\def\RR{{\mathbb R}}

\def\CC{{\mathbb C}}

\def\Re{{\rm Re \,}}
\def\Im{{\rm Im \,}}

\def\half{{1\over 2}}
\def\p{\partial}

\def\a{\alpha}
\def\k{\kappa}

\def\no{\nonumber}

\newcommand{\cform}[1]{\,{\cal C}\!\left[\protect\begin{matrix}#1\protect\end{matrix}\right]}
\newcommand{\cformtri}[3]{\,{\cal C}\!\left[\protect\begin{matrix}#1\protect\end{matrix}\middle|\protect\begin{matrix}#2\protect\end{matrix}\middle|\protect\begin{matrix}#3\protect\end{matrix}\right]}

\newcommand{\mfp}{\mathfrak{p}}

\DeclareMathOperator*{\esumsym}{\hspace{0.5em}\sum\raisebox{-0.5em}{\makebox[0.2em]{$\scriptstyle \mathrm{E}$}}\hspace{0.3em}}
\newcommand{\esum}{\hspace{-0.5em}\esumsym}

\begin{document}
\starttext
\setcounter{footnote}{0}

\vskip 1in
\begin{center}

{\Large \bf Holomorphic subgraph reduction of\\[0.5em] higher-point modular graph forms}

\vskip 0.3in

{\large  Jan E. Gerken$^1$ and Justin Kaidi$^2$} 

\vskip 0.3in
 { \sl ${}^1$ Max-Planck-Institut f{\"u}r Gravitationsphysik (Albert-Einstein-Institut)}\\
{\sl Am M{\"u}hlenberg 1, DE-14476 Potsdam, Germany }

\vskip 0.3in
 { \sl ${}^2$ Mani L. Bhaumik Institute for Theoretical Physics}\\
{\sl Department of Physics and Astronomy }\\
{\sl University of California, Los Angeles, CA 90095, USA} 

\vskip 0.2in

{\tt \small jan.gerken@aei.mpg.de, jkaidi@physics.ucla.edu}

\end{center}
\vskip 0.4in
\begin{abstract}
Modular graph forms are a class of modular covariant functions which appear in the genus-one contribution to the low-energy expansion of closed string scattering amplitudes. Modular graph forms with holomorphic subgraphs enjoy the simplifying property that they may be reduced to sums of products of modular graph forms of strictly lower loop order.  In the particular case of dihedral modular graph forms, a closed form expression for this holomorphic subgraph reduction was obtained previously by D'Hoker and Green. In the current work, we extend these results to trihedral modular graph forms. Doing so involves the identification of a modular covariant regularization scheme for certain conditionally convergent sums over discrete momenta, with some elements of the sum being excluded. The appropriate regularization scheme is identified for any number of exclusions, which in principle allows one to perform holomorphic subgraph reduction of higher-point modular graph forms with arbitrary holomorphic subgraphs.
\end{abstract}

\newpage


\newpage

\baselineskip=15pt
\setcounter{footnote}{0}

\tableofcontents
\newpage

\section{Introduction}
\label{sec:1}
In the genus-one contribution to the low-energy expansion of closed string amplitudes, a natural generalization of non-holomorphic Eisenstein series known as \textit{modular graph forms} arises. A modular graph form can be understood as an assignment of a certain modular covariant function\footnote{Throughout, we will call a function $f$ \emph{modular covariant of weight $(w,\bar w)$} if it transforms under modular transformations as
\begin{align}
  f\left(\frac{a\tau+b}{c\tau+d}\right)=(c\tau+d)^{w}(c\bar\tau+d)^{\bar w}f(\tau) \hspace{0.7 in}\begin{pmatrix}a&b\\c&d\end{pmatrix}\in\mathrm{SL}(2,\mathbb{Z})\no
\end{align}
We refer to $w$ and $\bar w$ as the holomorphic and anti-holomorphic weight, respectively.} to a scalar Feynman graph on the torus. As will be reviewed in more detail below, a general  modular graph form can be written as 
\begin{align}
\label{modgraphform1}
\mathcal{C}_{\Gamma}\! \left[ \begin{matrix} A \\ B \end{matrix} \right]  (\tau) =  \sum_{p_1,\dots,p_n \in \Lambda}'\, \prod_{r =1}^n { (\tau_2 / \pi)^{\half a_r + \half b_r} \over (p_r) ^{a_r} \, (\bar p _r) ^{b_r} }\, \prod_{i =1}^m \delta \left (\sum_{r =1}^n \Gamma _{i \, r} \, p_r \right)
\end{align}
 The index  $i=1,\cdots, m $ runs over all the vertices of the Feynman graph $\Gamma$, while the index $r=1,\cdots, n$ runs over all edges. The variables $p_r=m_r + n_r \tau$ take values in an integer lattice $\Lambda$ and may be interpreted as the discrete momenta along each edge of the graph; $\tau= \tau_1 + i \tau_2$  is the modular parameter of the torus. Throughout this work, a prime on a sum indicates that the point $p_r=0$ is excluded (in addition to any exclusions which are explicitly indicated). All of the information about the graph is contained in the connectivity matrix $\Gamma_{i r}$, which enforces momentum conservation at each vertex, as well as in the arrays
\begin{align}
A &= [a_1, \dots, a_n] &B&= [b_1, \dots, b_n]
\end{align}
which catalogue the exponents of the holomorphic momenta $p_{r}$ and anti-holomorphic momenta $\bar p_{r}$, respectively. The weight of this modular graph form is $\sum_{r=1}^n\left({a_r-b_r \over 2}, {b_r-a_r\over2}\right)$, which in particular is always integer since the sum in \eqref{modgraphform1} vanishes by antisymmetry if $\sum_{r=1}^n(a_r + b_r)$ is odd. 

An interesting special class of modular graph forms are those with \textit{holomorphic subgraphs}, namely those whose graphs contain a closed subgraph with only holomorphic momenta $p_{r}$ along its edges. A holomorphic subgraph containing $n$ vertices of valence greater than or equal to $2$ is naturally referred to as an \textit{$n$-point holomorphic subgraph}. By definition, the anti-holomorphic exponents $b_i$ vanish along the edges of a holomorphic subgraph, and hence the presence of such subgraphs is (depending on the graph topology) easily diagnosed by the presence of two or more zeros in the lower entries of the exponent matrix $\left[A \,\,\,B \right]^T$ on the left hand side of (\ref{modgraphform1}).

Modular graph forms with holomorphic subgraphs admit a reduction to sums of products of simpler modular graph forms. In other words, modular graph forms with holomorphic subgraphs are always reducible to more primitive components. The precise rules for such a reduction were first analyzed in \cite{DHoker:2016mwo}, and will be reviewed in Section \ref{sec:diholsubred}. However, those results were limited to the case of dihedral modular graph forms (i.e.\,forms whose graphs contain two vertices, to each of which at least three edges are attached). In the current work, we aim to extend these results to modular graph forms with higher number of vertices.

That such an extension is physically useful can be seen in recent works on one-loop four-gluon scattering in heterotic string theory \cite{upcoming2}. We will preview some of these applications in Section \ref{examplessec}. More broadly, since many other amplitudes in string theory, including graviton scattering in type IIB  with more than four external particles \cite{Green:2013bza,Broedel:2014vla}, as well as amplitudes involving fermionic particles \cite{Lee:2017ujn}, can be expanded using modular graph forms, it is to be expected that simplifying relations such as the ones derived in this paper will play an important role in many other contexts.

The derivation of holomorphic subgraph reduction formulae for higher-point modular graph forms is similar in spirit to the derivation in the dihedral case. However, there is one conceptual novelty which arises: in the process of obtaining holomorphic subgraph reduction formulae, one encounters divergent sums of the form
\begin{align}
\label{Q1def}
  \sum'_{p\neq p_1,\dots, p_n} {1 \over p}
\end{align}
where $n+1$ is the order of the holomorphic subgraph. These have to be replaced by suitably defined expressions to be denoted by $Q_1(p_1,\dots,p_n)$, therefore regularizing \eqref{Q1def}. For the case of $n=1$, the correct expression for \eqref{Q1def} in the context of holomorphic subgraph reduction was obtained in \cite{DHoker:2016mwo}. Generalizing the result of \cite{DHoker:2016mwo}, we make an ansatz with one free parameter, which is fixed by requiring that the final result be modular covariant. This leads to the expression
\begin{align}
\label{finalresult}
 Q_1(p_1,\dots,p_n) = -\sum_{i=1}^n {1 \over p_i} - {\pi \over (n+1) \tau_2}\sum_{i=1}^n(p_i - \bar p_i )
\end{align}
We finally show that the regularization guessed in this way coincides with the use of the Eisenstein summation prescription to evaluate the original sum. Since no ambiguity arises in the latter procedure, using \eqref{finalresult} for \eqref{Q1def} during holomorphic subgraph reduction produces correct identities. The benefit of doing things in this roundabout way is that, since Eisenstein summation is not invariant under lattice-shifts (see Appendix~\ref{tworegs} for details), it has to be applied carefully. In contrast, the regularization \eqref{finalresult} can be used in a more naive manner and is therefore more practical for computations.

This paper is structured as follows. We begin in Section \ref{sec:review} with a brief review of the origin of modular graph forms in physics and proceed in Section \ref{sec:diholsubred} to give a more technical overview of previous results on holomorphic subgraph reduction of dihedral modular graph forms. In Section \ref{sec:triholsubred}, we extend these previous results to trihedral modular graph forms. Here, we also discuss applications of the resulting formula to heterotic string amplitudes. Finally, in Section \ref{sec:regularization} we discuss the most general case of holomorphic subgraph reduction and prove that \eqref{finalresult} is the appropriate regularization of \eqref{Q1def} leading to modular covariant holomorphic subgraph reduction formulae.

\section{Modular graph functions and forms in physics}
\label{sec:review}
In addition to the massless supergravity spectrum, string theory predicts an infinite tower of massive particles with masses of order $(\a')^{-1/2}$. Though the direct production of such particles seems unlikely in the near or distant future, one may hope to identify this stringy spectrum indirectly through the effective interactions it induces. Such effective interactions are weighted by factors of $\a'$, and hence appear as an expansion in increasing numbers of derivatives of supergravity fields. If we restrict to the sector of effective interactions of type IIB involving only the graviton (with the axio-dilaton taken to be constant), then we have an effective action of the form \cite{DHokerLectureNotes}
\begin{align}
S_{\text{eff}} = {1 \over \k^2} \int d^{10}x \sqrt{-g_E} \sum_{m=1}^\infty \sum_{n=0}^\infty (\a')^{m+n-1} c_{m,n}(\eta) D^{2n} R^m + \dots
\end{align}
In the above, $D^{2n} R^m$ represents schematically some contraction of $2 n$ factors of the Einstein-frame covariant derivative with $m$ factors of the Einstein-frame curvature tensor. The exponent of $\a'$ was chosen such that the Einstein-Hilbert term is of order $O({\a'}^{\,0})$. The coefficient functions $c_{m,n}(\eta)$ are functions of the complex axio-dilaton  $\eta = \chi + i e^{- \phi}$, where we have avoided using the usual notation $\tau$ for this quantity since $\tau$ will be used to refer to the modular parameter of a torus in what follows. 

To what degree can the functions $c_{m,n}(\eta)$ be determined? Since $D^{2n} R^m$ is a scalar, so too are the functions $c_{m,n}(\eta)$.  Furthermore, because the action and Einstein-frame metric are invariant under $SL(2, \RR)$ (or rather $SL(2,\ZZ)$ in the full string theory), we expect that $c_{m,n}(\eta)$ enjoys this same property. However, what can be said beyond these simple results?

The most immediate set of further results follow from a linearized supersymmetry analysis, which reveals that besides the Einstein-Hilbert term,
\begin{align}
D^{2 n} R^m&= 0 &\mathrm{for}\,\,\,\,m&= 1,2,3
\end{align}
This statement can be recast as a prediction for superstring perturbation theory, namely that the one-, two-, and three-graviton amplitudes vanish. Indeed, this vanishing is well-known. Thus one may focus on the case of $m\geq 4$. In the simplest case of $m=4$, it is known \cite{Green:1997di,Green:1997as, Green:1999pu,Green:1999pv} that 
\begin{align}
\label{nonpertcoeffs}
c_{4,0}(\eta) &= \pi^{3/2}\, {\rm E}_{3/2}(\eta) & c_{4,1}(\eta)&= 0 &  c_{4,2}(\eta)&= \pi^{5/2}\, {\rm E}_{5/2}(\eta)
\end{align}
where $E_s(\eta)$ is the non-holomorphic Eisenstein series, defined as
\begin{align}
{\rm E}_s(\eta) = \sum'_{(m,n) \in \ZZ^2}{\eta_2^s \over \pi^s |m + n\, \eta|^{2s}}
\end{align}
where the prime superscript over the summation represents exclusion of the point $(m,n)=(0,0)$. These admit the following Fourier expansion,
\begin{align}
\label{nonholoE}
{\rm E}_s(\eta) &= 2  \pi^{-s}\, \eta_2^s \,\zeta(2s) + 2 \, {\Gamma\left(s-\half\right) \over \Gamma(s)} \pi^{\half - s} \,\eta_2^{1-s}\,\zeta(2s-1) 
\no\\
&\hphantom{=}+ {4 \sqrt{\eta_2} \over \Gamma(s)}\sum_{N \neq 0}N^{\half - s} \sigma_{2s-1}(|N|)\, e^{2 \pi i N \eta_1} \,K_{s-\half}(2 \pi \eta_2 |N|)
\end{align}
where $\zeta$ is the Riemann zeta function, $\sigma$ is the divisor function, and $K$ is the modified Bessel function of the second kind.

What is remarkable about the identifications in (\ref{nonpertcoeffs}) is that they are fully non-perturbative results. The perturbative contributions are given by the first two terms of (\ref{nonholoE}) which involve only $\eta_2=\Im(\eta)$, while all of the non-perturbative contributions are contained in the Fourier expansion involving the axion $\eta_1=\Re(\eta)$, which couples to D-instantons. From this, one concludes for example that the coefficient of the effective interaction $R^4$ receives perturbative corrections only at tree and one-loop level, while the  $D^4 R^4$ term receives perturbative corrections only at tree and two-loop level.\footnote{Recall that (\ref{nonholoE}) is given in Einstein frame, so that before counting powers of $\eta_2$ to determine the order in perturbation theory, we must multiply by $\eta_2^{1/2}$ to convert to string frame.} The vanishing of the two-loop contribution to $R^4$ was verified in \cite{DHoker:2005vch}, while the vanishing of the one-loop contribution and calculation of the two loop contribution to $D^4 R^4$ was performed in \cite{DHoker:2005jhf}. In addition, in \cite{Green:1997tv} the coupling of the axion to D-instantons was calculated and found to match with the non-perturbative portions of $c_{4,0}(\eta)$ in (\ref{nonholoE}).

Unfortunately, for $n>2$ there are far fewer non-perturbative results for the coefficients $c_{4,n}(\eta)$ (see \cite{DHoker:2014oxd} for a further review of what is known). Instead, in these cases one must generally settle for perturbative results obtained via calculation of four-graviton scattering amplitudes. For example, one may begin with the four-graviton tree-level amplitude $\cA_0^{(4)} $, which takes the familiar form 
\begin{align}
\label{treelevelA}
\cA_0^{(4)} &= {R^4 \eta_2^2 \over s t u} {\Gamma(1-s) \Gamma(1-t) \Gamma(1-u) \over \Gamma(1+s) \Gamma(1+t) \Gamma(1+u)}
\no\\
&= R^4 \eta_2^2 \left[{1 \over st u} + 2 \zeta(3) + \zeta(5) (s^2 + t^2 + u^2)+ 2 \zeta(3)^2 stu + O(s_{ij}^4) \right] 
\end{align}
where the Mandelstam invariants $s,t, u$ are given in terms of the Lorentz invariant combinations $s_{ij} = -{\a' \over 4}(k_i+k_j)^2$ via $ s= s_{12} = s_{34}$, $t=s_{14}=s_{23}$, and $u=s_{13}=s_{24}$. Note that in the above, we have used the fact that momentum conservation and the on-shell condition $k_i^2 = 0$ require $s+t+u=0$. The first term in (\ref{treelevelA}) is one-particle reducible and hence does not contribute to the $\alpha'$ corrections of the effective action, whereas each of the latter terms $O(s_{ij}^n)$ has an interpretation as the leading order contribution to the coefficient $c_{4,n}(\eta)$. From this point of view, the vanishing of (the first term of) $ c_{4,1}(\eta)$ is seen to be a result of the condition $s+t+u=0$.

Having identified the tree-level contributions to $c_{4,n}(\eta)$, we may now proceed to one-loop. In this case, the amplitude is given by \cite{DHoker:2015gmr}
\begin{align}
\cA_1^{(4)} = 2 \pi R^4 \int_{\cF} {d \tau_1 \wedge d\tau_2 \over \tau_2^2} \cB_4(s,t,u; \tau)
\end{align}
where $\tau$ is the modular parameter of the worldsheet torus and $\cF$ is the usual fundamental domain $\cF = \{\tau\in\CC\,\, \big| \,\, |\tau_1|\leq \half, \,\tau_2 > 0,\, |\tau| \geq 1 \}$. The partial amplitudes $\cB_4(s,t,u; \tau)$ may be written in terms of scalar Green's functions on the torus as
\begin{align}
\cB_4(s,t,u; \tau) = \prod_{i=1}^4 \int_\Sigma {d^2 z_i\over \tau_2}\, \mathrm{exp} \left\{ \sum_{1 \leq i < j \leq 4} s_{ij} G(z_i - z_j| \tau)\right\}\label{eq:8}
\end{align}
where $z_i$ are coordinates on the torus $\Sigma$.  The scalar Green's function on the torus admits the following Fourier representation, 
\begin{align}
\label{Greensfunctdef}
G(z|\tau) = \sum _{p \in \Lambda}' { \tau _2 \over \pi |p|^2 } \, e^{2 \pi i (n \alpha - m \beta)}
\end{align}
where $z=\alpha + \beta \tau$ with $\alpha, \beta \in \RR/\ZZ$.  The integers $m,n$ parametrize the discrete momenta of the torus $p=m + n \tau$, which take values in an integer lattice $\Lambda$. It is then clear that the partial amplitude $\cB_4(s,t,u; \tau)$ is a modular function of $\tau$.

We may now expand $\cB_4(s,t,u; \tau)$ in a power series in $s_{ij}$ to obtain 
\begin{align}
\label{Bpowerseries}
\cB_4(s,t,u; \tau) = \sum_{w=0}^\infty {1 \over w!}  \prod_{i=1}^4\int_\Sigma {d^2 z_i\over \tau_2} \,V(s,t,u; z_i; \tau)^w
\end{align}
where 
\begin{align}
V(s,t,u; z_i; \tau)= \sum_{1 \leq i < j \leq 4}s_{ij}  G(z_i - z_j| \tau)
\end{align}
In this form, we see that the term of order $w$ will contribute to the $D^{2w} R^4$ term in the effective action. Thus for example if $w=0$, then one finds 
\begin{align}
\cB_4(s,t,u; \tau) = \left( \int \frac{d^2 z_i}{\tau_{2}}\right)^4 = 1
\end{align}
and therefore
\begin{align}
\cA_1^{(4)} = 2 \pi R^4 \int_{\cF} {d \tau_1 \wedge d\tau_2 \over \tau_2^2}  = {2\pi^2 \over 3}R^4
\end{align}
Indeed, switching to Einstein frame, this means that the one-loop contribution to $c_{4,0}(\eta)$ is $ {2\pi^2 \over 3} \eta_2^{-1/2}$, which is exactly what is observed from (\ref{nonpertcoeffs}) and (\ref{nonholoE}).

For $w\neq0$, the partial amplitudes $\cB_4(s,t,u; \tau) $ are non-constant functions of $\tau$. It is clear that, by worldsheet modular invariance, these functions must be invariant under $SL(2,\ZZ)$ transformations. The functions arising in this context are known as \textit{modular graph functions} \cite{Green:2008uj,DHoker:2015gmr,DHoker:2015wxz}, and have been the subject of recent study in both physics \cite{DHoker:2016mwo,DHoker:2016quv,Basu:2016xrt,Basu:2016kli,Basu:2016mmk,Basu:2017nhs,Kleinschmidt:2017ege,Basu:2017zvt,Broedel:2018izr} and mathematics \cite{Brown:2017qwo,Brown:2017p2,DHoker:2017zhq,Zerbini:2018hgs,upcoming1}. They are special cases of the modular graph forms given in (\ref{modgraphform1}).

The name modular ``graph" function derives from the fact that these functions may be represented by Feynman graphs on the torus. As usual, we represent a Green's function graphically by an edge in a Feynman diagram,  
\begin{align}
\begin{tikzpicture}[baseline=-0.5ex,scale=1.7]
%
\draw (1,0) -- (2.5,0) ;
\draw (1,0) [fill=white] circle(0.05cm) ;
\draw (2.5,0) [fill=white] circle(0.05cm) ;
%
\draw (1,-0.25) node{$z_i$};
\draw (2.5,-0.25) node{$z_j$};
\end{tikzpicture}
  =~ G(z_i-z_j|\tau)
\label{fig1}
\end{align}
The integration over the position of a vertex $z$ on which $r$  Green's functions end is denoted by an unmarked  filled black dot, in contrast with an unintegrated vertex $z_i$ which is represented by a marked unfilled white dot. The basic ingredients in the graphical notation are depicted in the graph below,
\begin{align}
\begin{tikzpicture}[baseline=-0.5ex,scale=1.7]
%
\draw (2,0.7) -- (1,0) ;
\draw (2,0.7) -- (1.5,0) ;
\draw (2,0.7) -- (2.5,0) ;
\draw (2,0.7) -- (3,0) ;
\draw (2,0.2) node{$\cdots$};
\draw [fill=black]  (2,0.68)  circle [radius=.05] ;
\draw (1,0)    [fill=white] circle(0.05cm) ;
\draw (1.5,0) [fill=white] circle(0.05cm) ;
\draw (2.5,0)    [fill=white] circle(0.05cm) ;
\draw (3,0) [fill=white] circle(0.05cm) ;
%
\draw (1,-0.25) node{$z_1$};
\draw (1.5,-0.25) node{$z_2$};
\draw (2.57,-0.25) node{$z_{r-1}$};
\draw (3.1,-0.25) node{$z_r$};
\end{tikzpicture}
\label{fig2}
= \ \int _\Sigma {d^2 z \over  \tau_2} \, \prod _{i=1}^r G(z-z_i|\tau)
\end{align}
For our purposes, we will be interested only in those cases in which all positions on the torus have been integrated over, and hence all nodes in the diagram are filled and unmarked.

Thus far we have been discussing exclusively the four-graviton amplitudes, and the associated $D^{2n}R^4$ terms in the effective action. However, insofar as modular graph functions are concerned, we may easily generalize to functions corresponding to graphs with arbitrary numbers of vertices. If we denote by $\nu_{ij}$ the exponent of $s_{ij}$ in the expansion (\ref{Bpowerseries}), then the power series expansion of $\cB_m$  is given by a Feynman graph $\Gamma$ with associated integral, 
\begin{align}
\mathcal{C}_{\Gamma}(\tau) = \left ( \prod_{k=1}^m \int_{\Sigma} {{d^2 z_k}\over \tau_2} \right ) 
\prod_{1\leq i < j \leq m} G(z_i - z_j | \tau)^{\nu_{ij}}
\end{align}
The graph $\Gamma$ has $m$ vertices, labelled by $k=1,\cdots, m$ and $\nu_{ij}$ edges between vertices $i$ and $j$, with the total number of edges  given by the {\sl weight} $w$ of the graph $\Gamma$,\footnote{This notion of weight should not be confused with the \textit{modular weight} of the graph function, which in the current case is zero.} 
\begin{align}
w=\sum_{1 \leq i < j \leq m} \nu_{ij}
\end{align}
in analogy to the $w$ introduced in (\ref{Bpowerseries}). In terms of the Fourier series for the Green's function (\ref{Greensfunctdef}), this expression is given by, 
\begin{align}
\mathcal{C}_{\Gamma}(\tau) = \sum_{p_1,\ldots,p_w \in \Lambda}' 
\left ( \prod_{r=1}^w {{\tau_2}\over {\pi |p_{r}|^2}} \right ) 
\prod_{i=1}^m \delta\left(\sum_{r=1}^w \Gamma_{ir}p_{r}\right)\label{eq:26}
\end{align}
For such functions, all of the information about the graph $\Gamma$ is contained in its connectivity matrix $\Gamma_{i r}$, where the index $i=1,\cdots, m $ runs over all of the vertices of $\Gamma$ and the index $r=1,\cdots, w$ runs over all of the edges. When the edge $r$ does not end on the vertex $i$, we have $\Gamma _{i r}= 0$, while otherwise we have $\Gamma _{i r} = \pm 1$, with the sign depending on the orientation conventions for the momenta flowing into the vertices. 

Finally, one may further generalize this class of functions to modular graph \textit{forms}, which can have arbitrary exponents for their holomorphic and anti-holomorphic momenta. The general form of these was shown in (\ref{modgraphform1}). These objects are no longer modular invariant, but are manifestly modular covariant, transforming as
\begin{align}
\cC \left [ \begin{matrix}A \cr B \cr\end{matrix} \right ] \left  ({ \alpha \tau + \beta \over \gamma \tau + \delta} \right ) 
= \left ( { \gamma \tau+\delta \over \gamma  \bar \tau +\delta } \right ) ^{\frac{1}{2}(a-b)}
\cC \left [ \begin{matrix}A \cr B \cr\end{matrix} \right ] (\tau) 
\end{align}
where $\alpha, \beta , \gamma, \delta \in \ZZ$ and $\alpha \delta - \beta \gamma =1$.  
The total exponents of holomorphic and anti-holomorphic momenta are given respectively by the following sums,
\begin{align}
\label{ab}
a& = \sum _{r =1}^m  a_r & b&= \sum _{r =1} ^m b_r  
\end{align}
Note that the weights of modular graph forms are always integers since the sum in \eqref{modgraphform1} vanishes by antisymmetry if $a+b$ is odd.

Modular graph forms are an interesting generalization of modular graph functions that allow for the derivation of important relations between modular graph functions \cite{DHoker:2016mwo}. From the physical point of view, modular graph forms appear when the integrand in \eqref{eq:8} contains prefactors in front of the Koba-Nielsen factor \cite{upcoming2,Broedel:2014vla,Green:2013bza,Lee:2017ujn}. For a more detailed introduction to modular graph forms, see e.g. \cite{DHoker:2016mwo,DHoker:2016quv}. For recent extensions to higher genus, see \cite{DHoker:2017pvk,Basu:2018eep,DHoker:2018mys}.

\section{Holomorphic subgraph reduction of dihedral graphs}
\label{sec:diholsubred}
We now give a brief overview of the holomorphic subgraph reduction procedure for dihedral graphs, as introduced in \cite{DHoker:2016mwo}. Since we will present the calculation of the trihedral holomorphic subgraph reduction formulae in detail in Section \ref{sec:triholsubred}, we will refrain from providing technical details here, instead focusing on the main conceptual points that will carry over to the later calculation. For more details the reader is referred to \cite{DHoker:2016mwo}.

A generic dihedral modular graph form with a holomorphic subgraph may be represented by the following graph,
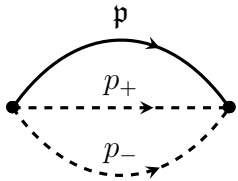
\begin{center}
  \begin{tikzpicture}[scale=1.2]
    %
    \draw[directed,very thick] (-1.2,0) node{$\bullet$}..controls(-0.5,1) and (0.5,1).. node[above]{$\mfp$}(1.2,0) node{$\bullet$};  
    \draw[directed,very thick,dashed] (-1.2,0) node{$\bullet$}-- node[above]{$p_{+}$}(1.2,0) node{$\bullet$};
    \draw[directed,very thick,dashed] (-1.2,0) node{$\bullet$}..controls(-0.5,-1) and (0.5,-1).. node[above]{$p_{-}$}(1.2,0) node{$\bullet$};  
  \end{tikzpicture}
\end{center}
The solid line represents an arbitrary number $\ell$ of parallel lines with total momentum $\mfp$ flowing from left to right. The dashed lines represent single lines with only holomorphic momenta flowing through them (i.e. vanishing exponents $b_\pm$, in our previous notation). The corresponding lattice sum is then given by
\begin{align}
  \cform{a_{+}&a_{-}&A\\0&0&B} =  \sum_{p_1,\dots,p_\ell, p_+,p_- \in \Lambda}'\,  { (\tau_2 / \pi)^{\half a_0} \over (p_+) ^{a_+}(p_-) ^{a_-}  }\, \prod_{r =1}^{\ell} { (\tau_2 / \pi)^{\half a_r + \half b_r} \over (p_r) ^{a_r} \, (\bar p _r) ^{b_r} }\,  \delta \left (\sum_{\a =0}^{\ell} \, p_\a \right)
\end{align}
where $a_0=a_{+}+a_{-}$, $p_0 = p_++ p_-$, and the exponents for each of the $\ell$ momenta in $\mfp$ are collected in $A,B$. In order for this sum to be absolutely convergent, we restrict to $a_{0}\geq3$. The basic strategy of holomorphic subgraph reduction is to isolate the two holomorphic edges, utilize the momentum-conserving delta-function to rewrite
\begin{align}
  \frac{1}{(p_{+})^{a_{+}}(p_{-})^{a_{-}}}=\frac{1}{(p_{+})^{a_{+}}(-\mfp-p_{+})^{a_{-}}}\label{eq:10}
\end{align}
and then to perform a partial-fraction decomposition in $p_+$. Once this has been done, the summation over $p_+$ can be performed explicitly. The resulting expression then has one less momentum, and thus one less edge, than the original modular graph form. This implies that modular graph forms with a holomorphic subgraph are reducible to sums of products of modular graph forms with fewer loops.

A subtlety in this procedure is that by naively distributing the sum over the partial fraction decomposition, conditionally convergent sums can be produced. In particular, sums of the form
\begin{align}
Q_{k}(p_{0})\equiv   \sum_{p\neq p_{0}}' \frac{1}{p^{k}}\hspace{0.7 in} k\geq1\label{eq:33}
\end{align}
arise, which are not absolutely convergent for $k=1,2$. To rectify this issue, we must find appropriate definitions for these sums. The definitions which were chosen in \cite{DHoker:2016mwo} are
\begin{align}
  Q_1(p_0) &= -{1 \over p_0} - {\pi \over 2 \tau_2} (p_0- \bar p_0) 
  \no\\
   Q_2(p_0)& = -{1 \over {p_0}^2} +\mathcal{\hat G}_2 + {\pi \over \tau_2}
  \no\\
  Q_k(p_0) &= - {1 \over {p_0}^k} + \cG_k \hspace{0.7 in} k \geq 3
  \label{oneexclusion}
\end{align}
Here, the functions $\mathcal{G}_{k}$ are defined as 
\begin{align}
  \mathcal{G}_{k}(\tau)&=\pi^{k/2}{\rm G}_{k}(\tau) \hspace{0.7 in} k\geq 3\\
  \mathcal{\hat G}_{2}(\tau)&=\pi {\rm \hat G}_{2}(\tau)  
\end{align}
 where the functions ${\rm G}_{k}$ are holomorphic Eisenstein series,\footnote{We warn the reader that our normalization of ${\rm G}_{k}$, which follows the conventions of \cite{DHoker:2016mwo,DHoker:2016quv}, is non-standard. The objects $\mathcal{G}_{k}$ have the more standard normalization, in which the $q$-expansion starts with a rational multiple of $\pi^k$.}
\begin{align}
  {\rm G}_{k}(\tau)=\sum_{(m,n)\neq (0,0)}\frac{1}{\pi^{k/2}(m\tau+n)^{k}}\hspace{0.5 in} k\geq 2
\end{align}
The function ${\rm \hat G}_{2}$ is the non-holomorphic but modular covariant regularization of the conditionally convergent series ${\rm G}_{2}$,
\begin{align}
  {\rm \hat G}_{2}(\tau)=\lim_{s\rightarrow 0}\sum_{(m,n) \neq (0,0)} \frac{1}{\pi(m\tau + n)^{2} \,|m\tau + n|^s}=\frac{\pi}{3}-8\pi\sum_{n=1}^{\infty}\sigma_{1}(n)q^{n}-\frac{1}{\tau_{2}}\label{eq:32}
\end{align}
with $\sigma_{k}(n)$ the divisor sum.

As will be explained in detail later, the choice \eqref{oneexclusion} is not unique. An important point is that the term $-\frac{\pi}{2\tau_{2}}p_{0}$ in $Q_{1}(p_{0})$ and the term $\frac{\pi}{\tau_{2}}$ in $Q_{2}(p_{0})$ have different modular weights than the sums on the respective left-hand sides. But when plugged into the full expression resulting from partial fraction decomposition of \eqref{eq:10}, these terms of  abnormal modular weight cancel out, leading to a total result with the expected modular properties. This will be a guiding principle for us in what follows. 

For completeness, we quote the final result for the holomorphic subgraph reduction of dihedral graphs forms \cite{DHoker:2016mwo},\footnote{There are some simple differences between our form of this equation and the form in \cite{DHoker:2016mwo}, but these are mostly due to conventions. One thing which is not due to convention, however, is the presence of an extra factor of $\pi \tau_2$ in the coefficient of the last term of (5.14) in  \cite{DHoker:2016mwo}, which should not be present.}
\begin{align}
  \cC \left [ \begin{matrix}a_+ & a_- & A \cr 0& 0 & B \cr\end{matrix} \right ] =&(-1)^{a_{-}}{\tau_{2}^{\frac{1}{2}a_{0}}}{\rm G}_{a_{0}}\cC \left [ \begin{matrix}A \cr B \cr\end{matrix} \right ] - \binom{a_0}{a_+}\,\cC \left [ \begin{matrix}a_0 & A \cr 0 & B \cr\end{matrix} \right ]
\no\\
  &+  \sum_{k=4}^{a_+} \binom{a_0-1-k}{a_+-k} \,{\tau_{2}^{\frac{1}{2}k}} {\rm G}_{k}  \cC \left[\begin{matrix}a_0-k & A \\ 0  & B\end{matrix} \right]
\no\\
                         &+ \sum_{k=4}^{a_-} \binom{a_0-1-k}{a_- -k}  \,{\tau_{2}^{\frac{1}{2}k}}{\rm G}_{k}  \cC \left[\begin{matrix}a_0-k & A \\ 0  & B\end{matrix} \right]
\no\\
                   &+\binom{a_0-2}{a_+-1} \left\{{\tau_{2}}{ \rm\hat G}_2 \,\cC \left[\begin{matrix}a_0-2 & A \\ 0  & B\end{matrix} \right]  + \cC \left[\begin{matrix}a_0-1 & A \\ -1  & B\end{matrix} \right]\right\}
    \label{eq:7}
\end{align}
\section{Extension to trihedral graphs}
\label{sec:triholsubred}
In this section, we will generalize the holomorphic subgraph reduction procedure outlined in the previous section to trihedral modular graph forms.

When the graph corresponding to a modular graph form has dihedral topology, it is sufficient to consider only two-point holomorphic subgraphs in order to arrive at a general formula for holomorphic subgraph reduction. For trihedral topology however, there are two cases that need to be distinguished: graphs with two- and three-point holomorphic subgraphs. The case of two-point holomorphic subgraphs will be treated in section \ref{sec:two-point-hsr}, and is a straightforward generalization of the dihedral result. We will therefore only quote the result for the decomposition formula in this case. In section \ref{sec:three-point-hsr}, we will discuss the case of three-point subgraphs. Since this case requires additional regularizations of the form \eqref{oneexclusion} and is considerably more complex, we derive the decomposition formula in full detail.

\subsection{Two-point holomorphic subgraph reduction}\label{sec:two-point-hsr}
A general trihedral graph with a two-point holomorphic subgraph is depicted in the following figure:
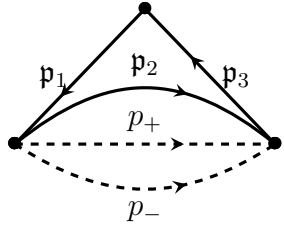
\begin{center}
  \begin{tikzpicture}[scale=1]
    \begin{scope}[xshift=-5cm,yshift=-0.4cm]
      \draw[directed,very thick]  (8.73,-1) node{$\bullet$} --node[right]{$\mfp_{3}$} (7,0.8) node{$\bullet$};
      \draw[directed,very thick]  (7,0.8)       node{$\bullet$} --node[left]{$\mfp_{1}$} (5.27,-1) node{$\bullet$};
      \draw[directed,very thick]  (5.27,-1) node{$\bullet$} .. controls (6.5,0) and (7.5,0) .. node[above]{$\mfp_{2}$}(8.73,-1) node{$\bullet$};
      \draw[directed,very thick,dashed] (5.27,-1) node{$\bullet$} --node[above]{$p_{+}$} (8.73,-1) node{$\bullet$};
      \draw[directed,very thick,dashed] (5.27,-1) node{$\bullet$} .. controls (6.5,-1.8) and (7.5,-1.8) .. node[below]{$p_{-}$}(8.73,-1) node{$\bullet$};
    \end{scope}
  \end{tikzpicture}
\end{center}
As before, solid lines  represent potentially several parallel edges whose momenta all flow in the indicated direction and add up to the momentum $\mfp_{i}$ in the label. Note that either $\mfp_{1}$ or $\mfp_{3}$ \textit{must} have more than one edge, lest the graph becomes dihedral.  The dashed lines, which represent single edges with holomorphic momenta, form a two-point holomorphic subgraph of the total graph. The general lattice sum for such graphs is
\begin{align}
  \label{target1}
  \cformtri{A_{1}\\B_{1}}{a_+ & a_- & A_2\\0 &0 & B_2}{A_{3}\\B_{3}}= \sum'_{p_j \in \Lambda}\left( \prod{1 \over \mfp^A \bar \mfp^B}\right){1 \over p_+^{a_+}p_-^{a_-}} \delta_{\mfp_1, \,\mfp_2+p_+ + p_-  }\delta_{\mfp_1 ,\,\mfp_3}
\end{align}
with the summation being over the momenta of each edge. The exponent arrays $A_i$, $B_i$, $i=1,2,3$, take the form 
\begin{align}
A_i = [ a_{1}^{(i)}, \dots, a_{R_i}^{(i)}] \hspace{1 in}B_i = [ b_{1}^{(i)}, \dots, b_{R_i}^{(i)}]
\end{align}
with $R_i$ giving the number of elements in the $i$-th array. As in the previous case we have introduced the collective momenta $\mfp_i$, defined by
\begin{align}
\mfp_i = \sum_{n_i=1}^{R_i} p_{n_i}^{(i)}
\end{align}
as well as the shorthand notation 
\begin{align}
\prod{1 \over \mfp^A \bar \mfp^B} \equiv \left(\frac{\tau_{2}}{\pi}\right)^{\frac{1}{2}(a_{+}+a_{-})}\prod_{i=1,2,3} \prod_{n_i=1}^{R_i} \left({\tau_2 \over \pi}\right)^{\half(a_{n_i}^{(i)}+b_{n_i}^{(i)})} {1 \over (p_{n_i}^{(i)})^{a_{n_i}^{(i)}}}{1 \over (\bar p_{n_i}^{(i)})^{b_{n_i}^{(i)}}}
\end{align}

 The formula for two-point trihedral holomorphic subgraph reduction is a straightforward generalization of that in the dihedral case. Because the holomorphic subgraph effectively forms a dihedral graph with the edges $(A_{2},B_{2})$, the result can be obtained from \eqref{eq:7} by replacing $[A,B]$ with $[A_{2},B_{2}]$ and adding the $[A_{1},B_{1}]$ and $[A_{3},B_{3}]$ blocks to the left and right in all $\mathcal{C}[\dots]$ expressions. For $a_{0}=a_{+}+a_{-}\geq3$, we have
 \begin{align}
  \cformtri{A_{1}\\B_{1}}{a_{+}&a_{-}&A_{2}\\0&0&B_{2}}{A_{3}\\B_{3}}=&(-1)^{a_{+}}{\tau_{2}^{\frac{1}{2}a_{0}}}{\rm G}_{a_{0}}\cformtri{A_{1}\\B_{1}}{A_{2}\\B_{2}}{A_{3}\\B_{3}}-\binom{a_{0}}{a_{+}}\cformtri{A_{1}\\B_{1}}{a_{0}&A_{2}\\0&B_{2}}{A_{3}\\B_{3}}\no\\
  &+\sum_{k=4}^{a_{+}}\binom{a_{0}-k-1}{a_{+}-k}{\tau_{2}^{\frac{1}{2}k}}{\rm G}_{k}\cformtri{A_{1}\\B_{1}}{a_{0}-k&A_{2}\\0&B_{2}}{A_{3}\\B_{3}}\nonumber\\
  &+\sum_{k=4}^{a_{-}}\binom{a_{0}-k-1}{a_{-}-k}{\tau_{2}^{\frac{1}{2}k}}{\rm G}_{k}\cformtri{A_{1}\\B_{1}}{a_{0}-k&A_{2}\\0&B_{2}}{A_{3}\\B_{3}}\no\\
    &+\binom{a_{0}-2}{a_{+}-1}\left({\tau_{2}}{\rm \hat G}_{2}\cformtri{A_{1}\\B_{1}}{a_{0}-2&A_{2}\\0&B_{2}}{A_{3}\\B_{3}}+\cformtri{A_{1}\\B_{1}}{a_{0}-1&A_{2}\\-1&B_{2}}{A_{3}\\B_{3}}\right)\label{trivalent2pt}
\end{align}

\subsection{Three-point holomorphic subgraph reduction}\label{sec:three-point-hsr}

We now proceed to the main focus of this work, which is holomorphic subgraph reduction of three-point holomorphic subgraphs in trihedral modular graph forms. 
The graphs in question are shown in the following figure,
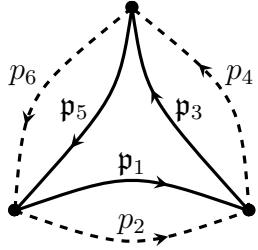
\begin{center}
  \begin{tikzpicture}[scale=0.9]
    \begin{scope}[xshift=-5cm,yshift=-0.4cm]
      \draw[directed,very thick,dashed]  (8.73,-1) node{$\bullet$} .. controls (8.6,0.7) .. (7,2) node{$\bullet$};
      \draw[directed,very thick]  (8.73,-1) node{$\bullet$} .. controls (7.2,0.7) .. (7,2) node{$\bullet$};
      \draw[directed,very thick, dashed]  (7,2)       node{$\bullet$} .. controls (5.4,0.7) ..(5.27,-1) node{$\bullet$};
      \draw[directed,very thick]  (7,2)       node{$\bullet$} .. controls (6.8,0.7) ..(5.27,-1) node{$\bullet$};
      \draw[directed,very thick,dashed] (5.27,-1) node{$\bullet$} .. controls (7,-1.6) .. (8.73,-1) node{$\bullet$};
      \draw[directed,very thick]  (5.27,-1) node{$\bullet$} .. controls (7,-0.4) .. (8.73,-1) node{$\bullet$};
      \draw (8.6,1) node{$p_4$};
      \draw (7-1.6,1) node{$p_6$};
      \draw (7+0,-0.3) node{$\mfp_1$};
      \draw (7+0,-1.2) node{$p_2$};
      \draw (7-0.85,0.7-0.2) node{$\mfp_5$};
      \draw (7+0.85,0.7-0.2) node{$\mfp_3$};
    \end{scope}
  \end{tikzpicture}
\end{center}
The dashed holomorphic edges form a three-point subgraph, and the general formula for such graphs is
\begin{align}
  \label{target}
  \cformtri{A_{1}&a_{2}\\B_{1}&0}{A_{3}&a_{4}\\B_{3}&0}{A_{5}&a_{6}\\B_{5}&0}= \sum'_{p_j \in \Lambda}\left( \prod{1 \over \mfp^A \bar \mfp^B}\right){1 \over p_2^{a_2}p_4^{a_4}p_6^{a_6}} \delta_{\mfp_1 + p_2,\mfp_3+ p_4 }\delta_{\mfp_3+ p_4 ,\mfp_5+ p_6 }
\end{align}
with the summation being over the momenta of each edge. The notation is as before, though we have redefined
\begin{align}
\prod{1 \over \mfp^A \bar \mfp^B} \equiv \left(\frac{\tau_{2}}{\pi}\right)^{\frac{1}{2}(a_{2}+a_{4}+a_{6})}\prod_{i=1,3,5} \prod_{n_i=1}^{R_i} \left({\tau_2 \over \pi}\right)^{\half(a_{n_i}^{(i)}+b_{n_i}^{(i)})} {1 \over (p_{n_i}^{(i)})^{a_{n_i}^{(i)}}}{1 \over (\bar p_{n_i}^{(i)})^{b_{n_i}^{(i)}}}
\end{align}
In what follows, we will also use the notation $\mfp_{ij} = \mfp_i - \mfp_j$ and $a_0 = a_2 + a_4 + a_6$. To evaluate (\ref{target}), we may begin by using the delta functions to replace $p_2$ and $p_4$ by $p_6$ and the various external momenta $\mfp_i$. In particular, we may rewrite 
\begin{align}
\label{target2}
\cformtri{A_{1}&a_{2}\\B_{1}&0}{A_{3}&a_{4}\\B_{3}&0}{A_{5}&a_{6}\\B_{5}&0}= \sum'_{p_{n}^{(i)}}\sum'_{p_6 \neq \mfp_{15}, \mfp_{35} } \left(\prod{1 \over \mfp^A \bar \mfp^B}\right){1 \over p_6^{a_6}(p_6-\mfp_{15})^{a_2}(p_6-\mfp_{35})^{a_4}}
\end{align}
Since $\prod{1 \over \mfp^A \bar \mfp^B}$ does not depend on $p_{6}$, we can focus on evaluating the following sum 
\begin{align}
\label{sumS}
\mathcal{S}= \sum'_{p_6 \neq \mfp_{15}, \mfp_{35} }{1 \over p_6^{a_6}(p_6-\mfp_{15})^{a_2}(p_6-\mfp_{35})^{a_4}}
\end{align}

\subsubsection{Decomposing $\mathcal{S}$}

In order to perform the sum \eqref{sumS}, we first separate out all cases in which $\mfp_{15}$ and $\mfp_{35}$ are equal to each other or to zero. In particular, there are five cases to study, 
\begin{align}
&\mfp_{15}= \mfp_{35}=0 & \mathcal{L}_1 &= \sum'_{p_6} {1 \over p_6^{a_0}}
\no\\
&\mfp_{15}= \mfp_{35}\neq0 &\mathcal{L}_2 &= \sum'_{p_6 \neq \mfp_{15}} {1 \over p_6^{a_6} (p_6 - \mfp_{15})^{a_2 + a_4}}
\no\\
&\mfp_{15} \neq 0 \,\,\,\,\mfp_{35}=0 &\mathcal{L}_3 &= \sum'_{p_6 \neq \mfp_{15}} {1 \over p_6^{a_6+a_4} (p_6 - \mfp_{15})^{a_2 }}
\no\\
&\mfp_{15} = 0 \,\,\,\,\mfp_{35}\neq 0 &\mathcal{L}_4&= \sum'_{p_6 \neq \mfp_{35}} {1 \over p_6^{a_6+a_2} (p_6 - \mfp_{35})^{a_4}}
\no\\
&\mfp_{15},\mfp_{35}\neq 0 \,\,\,\, \mfp_{15}\neq\mfp_{35}  & \mathcal{L}_5& =  \sum'_{p_6 \neq \mfp_{15}, \mfp_{35} }{1 \over p_6^{a_6}(p_6-\mfp_{15})^{a_2}(p_6-\mfp_{35})^{a_4}}  \label{sumsLi}
\end{align}
The function $\mathcal{S}$ is the sum of the above five terms. We may now evaluate them one by one. The first sum is trivial,
\begin{align}
\mathcal{L}_1 = {\mathcal G}_{a_0}
\end{align}
To evaluate the second sum, we begin by utilizing the following partial fraction identity
\begin{align}
\label{partfracdecom}
  \frac{1}{p^{a}(q-p)^{b}}=\sum_{k=1}^{a}\binom{a+b-k-1}{a-k}\frac{1}{p^{k}q^{a+b-k}}+\sum_{k=1}^{b}\binom{a+b-k-1}{b-k}\frac{1}{q^{a+b-k}(q-p)^{k}}
\end{align}
which allows us to rewrite $\mathcal{L}_2$ as 
\begin{align}
(-1)^{a_2 + a_4}\mathcal{L}_2 &= \sum'_{p_6 \neq \mfp_{15}}\left[\sum_{k=1}^{a_6} \binom{a_0-k-1}{a_6-k}{1 \over p_6^k\, \mfp_{15}^{a_0-k}} + \sum_{k=1}^{a_2+a_4} \binom{a_0-k-1}{a_2+a_4-k}{1 \over (\mfp_{15}-p_6)^k \mfp_{15}^{a_0-k}} \right]
\no\\
&= \sum_{k=1}^{a_6} \binom{a_0-k-1}{a_6-k}{Q_k(\mfp_{15}) \over \mfp_{15}^{a_0-k}}  + \sum_{k=1}^{a_2+a_4} \binom{a_0-k-1}{a_2+a_4-k}{Q_k(\mfp_{15}) \over  \mfp_{15}^{a_0-k}}
\end{align}
We now use the regularizations \eqref{oneexclusion} for the $Q_{k}$. Upon applying the following identities, 
\begin{align}
  \sum_{k=1}^{a_1}\binom{a_1+a_2-k-1}{a_1-k}+\sum_{k=1}^{a_2}\binom{a_1+a_2-k-1}{a_2-k}&= \binom{a_1+a_2}{a_1}\no\\
  \binom{a_{0}-3}{a_{2}+a_{4}-2}+\binom{a_{0}-3}{a_{6}-2}&=\binom{a_{0}-2}{a_{6}-1}
\end{align}
 the sum $\mathcal{L}_{2}$ simplifies to
\begin{align}
(-1)^{a_2 + a_4}\mathcal{L}_2 =& \sum_{k=4}^{a_6}\binom{a_0 - k- 1}{a_6-k} {{\mathcal G}_k \over \mfp_{15}^{a_0-k}}+ \sum_{k=4}^{a_2+a_4}\binom{a_0 - k- 1}{a_2+a_4-k} {{\mathcal G}_k \over \mfp_{15}^{a_0-k}}
\no\\
&\hspace{0.2 in}-\binom{a_0}{a_6}{1 \over \mfp_{15}^{a_0}} + \binom{a_0-2}{a_6 -1} {1 \over \mfp_{15}^{a_0-1}}\left(\mfp_{15} {\mathcal{ \hat G}}_2 + {\frac{\pi}{\tau_{2}}}\bar \mfp_{15} \right)\label{L2result}
\end{align}
Crucially, note that the ${\pi \over \tau_2}$ terms in $Q_2(p_1)$ have cancelled with the $ {\pi \over 2 \tau_2} p_0$ terms of $Q_1(p_1)$, just as in the dihedral case. Recall that this was necessary for obtaining a modular covariant final result, since such terms had different modular weight than the other terms. In what follows, we will need the generalization of this modular covariant regularization to sums with more exclusions, i.e. $Q_1(p_1, \dots, p_n)$. This is explored in detail in section \ref{sec:regularization}.

The sum $\mathcal{L}_3$ can be obtained from \eqref{L2result} by replacing $a_{6}\rightarrow a_{4}+a_{6}$ and $a_{2}+a_{4}\rightarrow a_{2}$. $\mathcal{L}_{4}$ can be reached by similar relabelings, so we may now proceed directly to $\mathcal{L}_5$. To begin, we apply the decomposition formula (\ref{partfracdecom}) twice to obtain
\begin{align}
\label{L5eq1}
(-1)^{a_2 + a_4} \mathcal{L}_5 =& \sum_{k=1}^{a_6}\sum_{\ell=1}^k \binom{a_2+a_6-k-1}{a_6-k}\binom{a_4+k-\ell-1}{k-\ell} {Q_\ell(\mfp_{15},\mfp_{35}) \over (\mfp_{15})^{a_2+a_6-k}(\mfp_{35})^{a_4+k-\ell}}
\no\\
& + \sum_{k=1}^{a_6}\sum_{\ell=1}^{a_4} \binom{a_2+a_6-k-1}{a_6-k}\binom{a_4+k-\ell-1}{a_4-\ell} {Q_\ell(\mfp_{31},\mfp_{35}) \over (\mfp_{15})^{a_2+a_6-k}(\mfp_{35})^{a_4+k-\ell}}
\no\\
& + \sum_{k=1}^{a_2}\sum_{\ell=1}^{a_4} \binom{a_2+a_6-k-1}{a_2-k}\binom{a_4+k-\ell-1}{a_4-\ell} (-1)^{k}{Q_\ell(\mfp_{31},\mfp_{35}) \over (\mfp_{15})^{a_2+a_6-k}(\mfp_{31})^{a_4+k-\ell}}
\no\\
&  +\sum_{k=1}^{a_2}\sum_{\ell=1}^{k} \binom{a_2+a_6-k-1}{a_2-k}\binom{a_4+k-\ell-1}{k-\ell} (-1)^{k}{Q_\ell(-\mfp_{15},\mfp_{31}) \over (\mfp_{15})^{a_2+a_6-k}(\mfp_{31})^{a_4+k-\ell}}
\end{align}
with the $Q_\ell(p_1,p_2)$ being the obvious generalization of (\ref{eq:33}) with two exclusions in the sum.

\subsubsection{Evaluating $\mathcal{L}_{5}$}
In order to evaluate \eqref{L5eq1}, we may insert the expressions for $Q_\ell(p_1,p_2)$ as in the case of $\mathcal{L}_2$. In particular, we use
\begin{align}
\label{twoexclusions}
  Q_{1}(p_1,p_2)&=-{1 \over p_1} - {1 \over p_2} - x {\pi \over \tau_2}(p_1 + p_2 - \bar p_1 - \bar p_2)
  \no\\
  Q_{2}(p_1,p_2)&=-{1 \over {p_1}^2}-{1 \over {p_2}^2} +\mathcal{\hat G}_2 + {\pi \over \tau_2}
  \no\\
  Q_k(p_1,p_2) &= -{1 \over {p_1}^k}-{1 \over {p_2}^k} +{\mathcal  G}_k \hspace{1 in} k\geq 3
\end{align}
These are the obvious generalizations of (\ref{oneexclusion}). The only subtlety is the choice of $x$ in the regularized expression for $ Q_{1}(p_1,p_2)$. As discussed before, we must choose $x$ such that all terms of abnormal modular weight cancel between $ Q_{1}(p_1,p_2)$ and $ Q_{2}(p_1,p_2)$. In fact, the correct choice is found to be $x=1/3$, and so we set $x$ equal to that value henceforth. This choice will be justified in Section \ref{sec:genreg}, where it arises as a special case of the general expression \eqref{eq:25} for an arbitrary number of excluded momenta.

Returning to the evaluation of $\mathcal{L}_5$, we may insert (\ref{twoexclusions}) and rewrite (\ref{L5eq1}) as 
\begin{align}
(-1)^{a_2 + a_4} \mathcal{L}_5 =&  \sum_{k=1}^{a_6} \binom{a_2+a_6-k-1}{a_6-k} {1 \over (\mfp_{15})^{a_2+a_6-k}(\mfp_{35})^{a_4+k}}\mathcal{X}_{k}( \mfp_{15},\mfp_{35}) 
\no\\
& + \sum_{k=1}^{a_2} \binom{a_2+a_6-k-1}{a_2-k} {(-1)^k \over (\mfp_{15})^{a_2+a_6-k}(\mfp_{31})^{a_4+k}}\mathcal{X}_{k}({-}\mfp_{15},\mfp_{31})\label{eq:2}
\end{align}
where we have defined the following function 
\begin{align}
\mathcal{X}_{k}(p, q) =& -\sum_{\ell=1}^k \binom{a_4+k-\ell-1}{k-\ell} \left(q \over p \right)^\ell - \sum_{\ell=1}^{a_4} \binom{a_4+k-\ell-1}{a_4-\ell} \left(q \over {q-p} \right)^\ell
\no\\
& + \sum_{\ell=4}^k \binom{a_4 + k -\ell - 1}{k - \ell} q^\ell {\mathcal G}_\ell + \sum_{\ell=4}^{a_4} \binom{a_4 + k -\ell - 1}{a_4 - \ell} q^\ell {\mathcal G}_\ell 
\no\\
&-\binom{a_4+k}{a_4} + \binom{a_4+k-2}{k-1}\left(q^2 \mathcal{\hat G}_2 + {\pi \over \tau_2} q \bar q \right)\label{eq:3}
\end{align}
 
\subsubsection{Summation over non-holomorphic momenta}

With \eqref{eq:3}, we have completed the evaluation of the five sums $\mathcal{L}_i$ listed in (\ref{sumsLi}) which make up the sum $\mathcal{S}$ in (\ref{sumS}). In order to obtain our final formula for three-point holomorphic subgraph reduction of (\ref{target2}), we must now carry out the sums over the remaining momenta. We denote the completely summed versions of the $\mathcal{L}_{i}$ by $L_{i}$, such that our final answer is given by
\begin{align}
  \cC \left [ \begin{matrix}A_1 \, a_2 \cr B_1 \, 0 \cr \end{matrix}\middle| \begin{matrix}A_3 \, a_4 \cr B_3 \, 0 \cr \end{matrix}\middle| \begin{matrix}A_5 \, a_6 \cr B_5 \, 0 \cr \end{matrix} \right ] = \sum_{i=1}^5 L_i
\end{align}
To begin, one calculates (recall that $a_{0}=a_{2}+a_{4}+a_{6}$)
\begin{align}
L_1 = \sum'_{p_n^{(i)}} \left(\prod{1 \over \mfp^A \bar \mfp^B}\right) \mathcal{L}_1 \,\delta_{\mfp_1,\mfp_3}\delta_{\mfp_3, \mfp_5} ={\tau_{2}^{\frac{1}{2}a_{0}}} {\rm G}_{a_0} \,\cformtri{A_1 \\B_1}{A_3 \\ B_3}{A_5\\B_5}\label{eq:5}
\end{align}
For the second contribution (\ref{L2result}), one has 
\begin{align}
    (-1)^{a_2 + a_4}L_2 =&\sum'_{\substack{p_{n}^{(i)}\\ \mfp_1 \neq \mfp_5}}\left(\prod{1 \over \mfp^A \bar \mfp^B}\right) (-1)^{a_2 + a_4} \mathcal{L}_2 \,\delta_{\mfp_1,\mfp_3}\no\\
=&  \sum_{k=4}^{a_6} \binom{a_0-k-1}{a_6-k}{\tau_{2}^{\frac{1}{2}k}} {\rm G}_k \,\cC\left [ \begin{matrix}A_1 \,  \cr B_1\, \cr \end{matrix}\middle| \begin{matrix}A_3 \cr B_3 \cr \end{matrix}\middle| \begin{matrix}\,A_5 & a_0-k \cr\, B_5 & 0\cr \end{matrix} \right ] 
\no\\
&+ \sum_{k=4}^{a_2+a_4} \binom{a_0-k-1}{a_2+a_4-k} {\tau_{2}^{\frac{1}{2}k}}{\rm G}_k \,\cC\left [ \begin{matrix}A_1 \,  \cr B_1\, \cr \end{matrix}\middle| \begin{matrix}A_3 \cr B_3 \cr \end{matrix}\middle| \begin{matrix}\,A_5 & a_0-k \cr\, B_5 & 0\cr \end{matrix} \right ]
\no\\
& + \binom{a_0-2}{a_6-1} \left\{{\tau_2} {\rm \hat  G}_2\,\cC\left [ \begin{matrix}A_1 \,  \cr B_1\, \cr \end{matrix}\middle| \begin{matrix}A_3 \cr B_3 \cr \end{matrix}\middle| \begin{matrix}\,A_5 & a_0-2 \cr\, B_5 & 0\cr \end{matrix} \right ] + \cC\left [ \begin{matrix}A_1 \,  \cr B_1\, \cr \end{matrix}\middle| \begin{matrix}A_3 \cr B_3 \cr \end{matrix}\middle| \begin{matrix}\,A_5 & a_0-1 \cr\, B_5 & -1\cr \end{matrix} \right ]    \right\}
\no\\
& - \binom{a_0}{a_6} \cC\left [ \begin{matrix}A_1 \,  \cr B_1\, \cr \end{matrix}\middle| \begin{matrix}A_3 \cr B_3 \cr \end{matrix}\middle| \begin{matrix}\,A_5 ~ a_0 \cr\, B_5 ~ 0\cr \end{matrix} \right ] \label{eq:1}
\end{align}
The third contribution can be obtained from this by relabeling $a_{6}\rightarrow a_{4}+a_{6}$ and $a_{2}+a_{4}\rightarrow a_{2}$, moving the $\mfp_{15}$-column to the first block and introducing an overall sign,
\bea
 (-1)^{a_4 + a_6} L_3 &=&  \sum_{k=4}^{a_4+a_6} \binom{a_0-k-1}{a_4+a_6-k} {\tau_{2}^{\frac{1}{2}k}} {\rm G}_k \,\cC\left [ \begin{matrix}A_1 & a_0-k \, \cr B_1 & 0\, \cr \end{matrix}\Bigg | \begin{matrix}A_3 \cr B_3 \cr \end{matrix}\Bigg | \begin{matrix}\,A_5  \cr\, B_5 \cr \end{matrix} \right ] 
\no\\
&\vphantom{.}&+ \sum_{k=4}^{a_2} \binom{a_0-k-1}{a_2-k} {\tau_{2}^{\frac{1}{2}k}} {\rm G}_k \,\cC\left [ \begin{matrix}A_1 & a_0-k \, \cr B_1 & 0 \,\cr \end{matrix}\Bigg | \begin{matrix}A_3 \cr B_3 \cr \end{matrix}\Bigg | \begin{matrix}\,A_5  \cr\, B_5 \cr \end{matrix} \right ] 
\no\\
&\vphantom{.}& + \binom{a_0-2}{a_4+a_6-1} \left\{ \tau_2 {\rm \hat  G}_2\,\cC\left [ \begin{matrix}A_1 & a_0-2\,  \cr B_1 & 0\, \cr \end{matrix}\Bigg | \begin{matrix}A_3 \cr B_3 \cr \end{matrix}\Bigg | \begin{matrix}\,A_5  \cr\, B_5 \cr \end{matrix} \right ] +\cC\left [ \begin{matrix}A_1 & a_0-1 \, \cr B_1 & -1\, \cr \end{matrix}\Bigg | \begin{matrix}A_3 \cr B_3 \cr \end{matrix}\Bigg | \begin{matrix}\,A_5  \cr\, B_5 \cr \end{matrix} \right ]     \right\}
\no\\
&\vphantom{.}& - \binom{a_0}{a_4+a_6} \cC\left [ \begin{matrix}A_1 ~ a_0  \cr B_1 ~ 0 \cr \end{matrix}\Bigg | \begin{matrix}A_3 \cr B_3 \cr \end{matrix}\Bigg | \begin{matrix}\,A_5  \cr\, B_5 \cr \end{matrix} \right ] 
\label{eq:4}
\eea
The fourth contribution can be obtained by relabeling $a_{6}\rightarrow a_{2}+a_{6}$ and $a_{2}+a_{4}\rightarrow a_{4}$ in \eqref{eq:1}, moving the $\mfp_{15}$-column to the second block and introducing an overall sign,
\bea
\label{cL4def}
(-1)^{a_2 + a_6} L_4 &=&  \sum_{k=4}^{a_2+a_6} \binom{a_0-k-1}{a_2+a_6-k} {\tau_{2}^{\frac{1}{2}k}}{\rm G}_k \,\cC\left [ \begin{matrix}A_1  \, \cr B_1 \, \cr \end{matrix}\Bigg | \begin{matrix}A_3 & a_0-k \cr B_3 & 0 \cr \end{matrix}\Bigg | \begin{matrix}\,A_5  \cr\, B_5 \cr \end{matrix} \right ] 
\no\\
&\vphantom{.}&+ \sum_{k=4}^{a_4} \binom{a_0-k-1}{a_4-k} {\tau_{2}^{\frac{1}{2}k}} {\rm G}_k \,\cC\left [ \begin{matrix}A_1  \, \cr B_1 \, \cr \end{matrix}\Bigg | \begin{matrix}A_3 & a_0-k \cr B_3 & 0 \cr \end{matrix}\Bigg | \begin{matrix}\,A_5  \cr\, B_5 \cr \end{matrix} \right ] 
\no\\
&\vphantom{.}& + \binom{a_0-2}{a_2+a_6-1} \left\{ \tau_2 {\rm \hat  G}_2\,\cC\left [ \begin{matrix}A_1 \, \cr B_1 \, \cr \end{matrix}\Bigg | \begin{matrix}A_3 & a_0-2 \cr B_3 & 0 \cr \end{matrix}\Bigg | \begin{matrix}\,A_5  \cr\, B_5 \cr \end{matrix} \right ]  +\cC\left [ \begin{matrix}A_1  \, \cr B_1 \, \cr \end{matrix}\Bigg | \begin{matrix}A_3 & a_0-1 \cr B_3 & -1 \cr \end{matrix}\Bigg | \begin{matrix}\,A_5  \cr\, B_5 \cr \end{matrix} \right ]     \right\}
\no\\
&\vphantom{.}& - \binom{a_0}{a_2+a_6}\cC\left [ \begin{matrix}A_1  \, \cr B_1 \, \cr \end{matrix}\Bigg | \begin{matrix}A_3 & a_0 \cr B_3 & 0 \cr \end{matrix}\Bigg | \begin{matrix}\,A_5  \cr\, B_5 \cr \end{matrix} \right ] 
\eea
Finally, we must consider the contribution due to $\mathcal{L}_{5}$, which is given by \eqref{eq:2} summed over the remaining momenta. To simplify the result, we introduce the following shorthand notation
\begin{align}
\cC\left [ \begin{matrix} m_1 \cr n_1 \, \cr \end{matrix}\middle| \begin{matrix}m_2 \cr  n_2 \cr \end{matrix}\middle|\,\,  \right ] &\equiv (-1)^{m_1+n_1+m_2+n_2} \,\cC\left [ \begin{matrix}A_1 ~m_1  \, \cr B_1 ~ n_1\, \cr \end{matrix}\middle| \begin{matrix}A_3 ~ m_2 \cr B_3 ~ n_2 \cr \end{matrix}\middle| \begin{matrix}\,A_5  \cr\, B_5 \cr \end{matrix} \right ] - \cC\left [ \begin{matrix}A_1  \, \cr B_1 \, \cr \end{matrix}\middle| \begin{matrix}A_3  \cr B_3 \cr \end{matrix}\middle| \begin{matrix}\,A_5~m_1+m_2  \cr\, B_5 ~ n_1+n_2 \cr \end{matrix} \right ]
\intertext{as well as }
\cC\left [ \,\, \middle| \begin{matrix} m_1 \cr n_1 \, \cr \end{matrix}\middle| \begin{matrix}m_2 \cr  n_2 \cr \end{matrix}  \right ] &\equiv (-1)^{m_2+n_2} \,\cC\left [ \begin{matrix}A_1   \, \cr B_1 \, \cr \end{matrix}\middle| \begin{matrix}A_3 ~ m_2 \cr B_3 ~ n_2 \cr \end{matrix}\middle| \begin{matrix}\,A_5 ~m_1  \cr\, B_5 ~n_1 \cr \end{matrix} \right ] -(-1)^{m_1+n_1} \,\cC\left [ \begin{matrix}A_1 ~m_1+m_2 \, \cr B_1 ~ n_1 +n_2 \, \cr \end{matrix}\middle| \begin{matrix}A_3  \cr B_3 \cr \end{matrix}\middle| \begin{matrix}\,A_5  \cr\, B_5  \cr \end{matrix} \right ]
\end{align}
Using the partial-fraction identity \eqref{partfracdecom} one final time to decompose the $(p-q)^{\ell}$ term in \eqref{eq:3}, we find the final result,
\begin{align}
\label{finalL5}
  (-1)^{a_2 + a_4}L_5 =&\sum_{k=1}^{a_6} \binom{a_2+a_6-k-1}{a_6-k} X_k{(0)} + \sum_{k=1}^{a_2} \binom{a_2+a_6-k-1}{a_2-k} (-1)^{k}\, \tilde{X}_k{(1)}
\end{align}
where we have defined
\begin{align}
\label{Ykdef}
X_k{(\epsilon)}\equiv& -\binom{a_4 + k}{a_4} \,\cC\left [ \begin{matrix}a_2+a_6-k \cr 0 \, \cr \end{matrix}\Big | \begin{matrix}a_4+k \cr  0 \cr \end{matrix}\Big |  \,\,\right ]
\no\\
& - \sum_{\ell=1}^k \binom{a_4 + k - \ell-1}{k-\ell}{(-1)^{\epsilon\ell}}\,\cC\left [ \begin{matrix}a_2+a_6-k+\ell \cr 0 \, \cr \end{matrix}\Big | \begin{matrix}a_4+k-\ell \cr  0 \cr \end{matrix}\Big |  \,\,\right ] 
\no\\
&+\sum_{\ell=4}^k \binom{a_4 + k - \ell-1}{k-\ell}{\tau_{2}^{\frac{1}{2}\ell}}{\rm G}_\ell \,\cC\left [ \begin{matrix}a_2+a_6-k \cr 0 \, \cr \end{matrix}\Big | \begin{matrix}a_4+k-\ell \cr  0 \cr \end{matrix}\Big |\,\,  \right ] 
\no\\
&+ \sum_{\ell=4}^{a_4} \binom{a_4 + k - \ell-1}{a_4-\ell} {\tau_{2}^{\frac{1}{2}\ell}}{\rm G}_\ell \,\cC\left [ \begin{matrix}a_2+a_6-k \cr 0 \, \cr \end{matrix}\Big | \begin{matrix}a_4+k-\ell \cr  0 \cr \end{matrix}\Big | \,\, \right ] 
\no\\
&+ \binom{a_4 +k-2}{k-1}\left\{{\tau_{2}} {\rm \hat G}_2 \, \cC\left [ \begin{matrix}a_2+a_6-k \cr 0 \, \cr \end{matrix}\Big | \begin{matrix}a_4+k-2 \cr  0 \cr \end{matrix}\Big |\,\,  \right ] + \cC\left [ \begin{matrix}a_2+a_6-k \cr 0 \, \cr \end{matrix}\Big | \begin{matrix}a_4+k-1 \cr  -1 \cr \end{matrix}\Big |\,\,  \right ]   \right\}
\no\\
&{-} \sum_{\ell=1}^{{a_{4}}} \binom{a_4 + k -\ell-1}{a_4 - \ell} (-1)^\ell \left\{\sum_{m=1}^{a_4 + k - \ell} \binom{a_4+k -m-1}{a_4 + k - \ell - m} (-1)^{\epsilon(a_{4}+k-m)}\cC\left [ \begin{matrix}a_0-m \cr 0 \, \cr \end{matrix}\Big | \begin{matrix}m \cr  0 \cr \end{matrix}\Big | \,\, \right ]\right.
\no\\
&\hphantom{- \sum_{\ell=1}^{a_{4}} \binom{a_4 + k -\ell-1}{a_6 - \ell} (-1)^\ell \Big\{}\left.+ \sum_{m=1}^{ \ell} \binom{a_4+k -m-1}{\ell - m}(-1)^{m}(-1)^{\epsilon(a_{4}+k-m)}\cC\left [\,\, \Big | \begin{matrix}a_0-m \cr 0 \, \cr \end{matrix}\Big|\begin{matrix}m \cr  0 \cr \end{matrix}  \right ] \right\}
\end{align}
and $\tilde{X}_k{(\epsilon)}$ is obtained from $X_k{(\epsilon)}$ by replacing all $\cC\left [ \begin{matrix} m_1 \cr n_1 \, \cr \end{matrix}\Big | \begin{matrix}m_2 \cr  n_2 \cr \end{matrix}\Big |\,\,  \right ] $ with $\cC\left [ \,\, \Big| \begin{matrix} m_1 \cr n_1 \, \cr \end{matrix}\Big | \begin{matrix}m_2 \cr  n_2 \cr \end{matrix}  \right ]$ and vice versa.

This completes the derivation of the three-point holomorphic subgraph reduction formula, which as stated above is given by
\begin{align}
    \cC \left [ \begin{matrix}A_1 \, a_2 \cr B_1 \, 0 \cr \end{matrix}\middle| \begin{matrix}A_3 \, a_4 \cr B_3 \, 0 \cr \end{matrix}\middle| \begin{matrix}A_5 \, a_6 \cr B_5 \, 0 \cr \end{matrix} \right ] = \sum_{i=1}^5 L_i\label{eq:9}
\end{align}
with the $L_i$ defined in equations (\ref{eq:5})-(\ref{cL4def}) and (\ref{finalL5}). Although this final result is rather lengthy, it is straightforward to implement it on a computer and provides simplifications for all trihedral graphs with three-point holomorphic subgraphs. 

\subsubsection{Divergent modular graph forms in the reduced expression}

When applying this formula one must be careful with the order in which the three blocks of the trihedral function are plugged into the formula, since an incorrect choice leads to divergent modular graph forms in the result. These divergences manifest themselves in
\begin{align}
  \begin{bmatrix}
    \,\,\,1&\,\,1\\
    -1&\,\,1
  \end{bmatrix}\label{eq:11}
\end{align}
subblocks appearing in the resulting modular graph forms. Naively using momentum conservation identities to simplify these leads to subblocks of the form
\begin{align}
  \begin{bmatrix}
    1&\,\,1\\
    0&\,\,0
  \end{bmatrix}
\end{align}
The sums corresponding to such modular graph forms are then divergent. Looking at the explicit expressions for the $L_{i}$ above (and recalling that $a_{0}\geq3$) shows that such $(1,-1)$ columns can only appear in the last term in the fifth line of \eqref{Ykdef} if $a_{4}=1$. In this case a $(1,-1)$ column is introduced in the second (middle) block of the modular graph form from both $X_{1}$ and $\tilde X_{1}$. This means that if $a_{4}=1$ and the $(A_{3},B_{3})$ block of the original modular graph form contains a $(1,1)$ column, divergent graphs will be produced by \eqref{eq:9}. Just like the divergence appearing upon partial fraction decomposition, this divergence is man-made - it results from an inappropriate application of the holomorphic subgraph reduction formula. 

There is an easy way to avoid this potential issue. From the original definition of trihedral modular graph forms, it is irrelevant in which order the three blocks of exponents are written. We may then rearrange the three blocks in such a way that the middle block does not contain a $\left[\begin{smallmatrix}1&1\\1&0\end{smallmatrix}\right]$-subblock.\footnote{The case in which all three blocks contain subblocks of the form $\left[\begin{smallmatrix}1&1\\1&0\end{smallmatrix}\right]$ cannot be reduced using \eqref{eq:9}.} This then avoids the problem altogether. We will see an explicit example of this below.

\subsection{Examples}
\label{examplessec}
We now offer a few examples to illustrate the utility of the three-point holomorphic subgraph reduction formula. First, consider the following trihedral modular graph form,
\begin{align}
  \cformtri{1&2\\1&0}{1&1\\1&0}{1\\0}=\sum'_{p_{i}\in\Lambda}\left(\frac{\tau_{2}}{\pi}\right)^{4}\frac{1}{p_{1}\bar p_{1}p_{3}\bar p_{3}}\frac{1}{p_{2}^{2}p_{4}p_{6}}\delta_{p_{1}+p_{2},p_{6}}\delta_{p_{3}+p_{4},p_{6}}
\end{align}
which appears in the calculation of four-gluon scattering in heterotic string theory at second order in $\alpha'$ \cite{upcoming2}. This modular graph form has a three-point holomorphic subgraph which may be reduced. However, in the current form the middle block is $\left[\begin{smallmatrix}1&1\\1&0\end{smallmatrix}\right]$, so naive application of the formulas above would produce divergent terms if  applied directly. To avoid this, we instead consider the equivalent expression
\begin{align}
  \cformtri{1&1\\1&0}{1&2\\1&0}{1\\0}=\sum'_{p_{i}\in\Lambda}\left(\frac{\tau_{2}}{\pi}\right)^{4}\frac{1}{p_{1}\bar p_{1}p_{3}\bar p_{3}}\frac{1}{p_{2}p_{4}^{2}p_{6}}\delta_{p_{1}+p_{2},p_{3}+p_{4}}\delta_{p_{1}+p_{2},p_{6}}
\end{align}
 Applying the results of section \ref{sec:three-point-hsr} to this yields for the $L_{i}$
\begin{align}
  L_{1}&=L_{3}=L_{4}=0\\
  L_{2}&=4 \cform{ 6 & 0 \\ 2 & 0 }-\cform{ 5 & 0 \\ 1 & 0 }-\tau_{2}{\rm \hat G}_{2}  \cform{ 4 & 0 \\ 2 & 0 }\\
  L_{5}&=-X_{1}+\tilde X_{1}
\end{align}
where
\begin{align}
  X_{1}&=3 \cform{ 6 & 0 \\ 2 & 0 }-\cform{ 5 & 0 \\ 1 & 0 }-\tau_{2}{\rm \hat G}_{2}  \cform{ 4 & 0 \\ 2 & 0 }-\cform{ 1 & 2 & 3 \\ 1 & 0 & 1 }\\
  \tilde X_{1}&=-\cform{ 6 & 0 \\ 2 & 0 }+\cform{ 3 & 0 \\ 1 & 0 }^2+\tau_{2} {\rm \hat G}_{2} \cform{ 1 & 1 & 2 \\ 0 & 1 & 1 }+\cform{ 1 & 2 & 2 \\ 1 & -1 & 1 }-3 \cform{ 1 & 2 & 3 \\ 1 & 1 & 0 }
\end{align}
Using momentum conservation and further straightforward identities between dihedral modular graph forms, one finds
\begin{align}
  \cformtri{1&1\\1&0}{1&2\\1&0}{1\\0}=&-\frac{1}{2} \cform{ 6 & 0 \\ 2 & 0 }+3 \cform{ 5 & 0 \\ 1 & 0 }+\frac{3}{2} \cform{ 3 & 0 \\ 1 & 0 }^2- \tau_{2} {\rm \hat G}_{2}\cform{ 3 & 0 \\ 1 & 0 }-\frac{1}{2}\tau_{2}{\rm \hat G}_{2} \cform{ 4 & 0 \\ 2 & 0 }-{\rm G}_{4} \tau_{2}^2\label{eq:12}
\end{align}

Another modular graph form which appears at second order in $\alpha'$ in the heterotic four-gluon scattering calculation is
\begin{align}
 \cC\left[ \begin{matrix} 2 \\ 0\end{matrix}   \middle| \begin{matrix} 1 & 1 \\ 1 & 0\end{matrix} \middle| \begin{matrix} 1 & 1 \\ 1 & 0\end{matrix} \right]= \sum'_{p_i \in \Lambda} \delta_{p_1,p_2+p_3} \delta_{p_1,p_4+ p_5} \left({\tau_2 \over \pi} \right)^{4}{1 \over p_2 \bar p_2 p_4 \bar p_4}\,{1 \over p_1^2 p_3 p_5} 
\end{align}
Using the three-point holomorphic subgraph reduction formula, this simplifies to
\begin{align}
  \cformtri{2\\0}{1&1\\1&0}{1&1\\1&0}&=2 \tau_2 {\rm \hat G}_{2} \cform{ 3 & 0 \\ 1 & 0 }+\tau _2{\rm \hat G}_{2}\cform{ 4 & 0 \\ 2 & 0 }-2 \cform{ 3 & 0 \\ 1 & 0 }^2-6 \cform{ 5 & 0 \\ 1 & 0 }+2 \cform{ 6 & 0 \\ 2 & 0 }+2 \tau _2^2 {\rm G}_{4}\label{eq:13}
\end{align}
Note that the decompositions \eqref{eq:12} and \eqref{eq:13} can be used to check our regularization scheme (\ref{twoexclusions}) with $x=1/3$. This is because these can be reduced to dihedral holomorphic subgraph reduction without introducing new divergent sums by doing a more careful partial fraction decomposition hand-tailored to these specific examples. A calculation along those lines is outlined in Appendix~\ref{sec:two-exampl-carefully}.

Finally, at third order in $\alpha'$ in the heterotic calculation, a more complex example arises which decomposes into dihedral graphs and lower-loop trihedral graphs,
\begin{align}
  \cC\left[ \begin{matrix} 2 & 1 \\ 1& 0\end{matrix}   \middle| \begin{matrix} 1 & 1 \\ 1 & 0\end{matrix} \middle| \begin{matrix} 1 & 1 \\ 1 & 0\end{matrix} \right]&=\,4 \cformtri{ 1 \\ 1 }{ 1 & 1 \\ 0 & 1 }{ 2 & 2 \\ 0 & 1 }+2 \cformtri{ 1 \\ 1 }{ 1 & 2 \\ 0 & 1 }{ 1 & 2 \\ 1 & 0 }-6  \tau_{2}{\rm \hat G}_{2} \cform{ 3 & 0 \\ 1 & 0 }-3 \tau_{2}{\rm \hat G}_{2} \cform{ 4 & 0 \\ 2 & 0 }\no\\
                             &\hphantom{=}-2\tau_{2} {\rm \hat G}_{2}  \cform{ 1 & 1 & 3 \\ 0 & 1 & 2 }+ \tau_{2}{\rm \hat G}_{2} \cform{ 1 & 2 & 2 \\ 0 & 1 & 2 }+6 \cform{ 3 & 0 \\ 1 & 0 }^2+18 \cform{ 5 & 0 \\ 1 & 0 }-4 \cform{ 6 & 0 \\ 2 & 0 }\no\\
  &\hphantom{=}-3 \cform{ 2 & 2 & 3 \\ 1 & 2 & 0 }-6 \tau_{2}^2{\rm G}_{4}\label{eq:21}
\end{align}
Note that to obtain this result, in addition to \eqref{eq:9} some other, more straightforward identities between modular graph forms were utilized.


\section{Definition of $Q_1(p_1,\dots,p_n)$}
\label{sec:regularization}

As we have seen in the previous section, our derivation of holomorphic subgraph reduction formulae relies on the regularization of sums of the form 
\begin{align}
 \sum'_{p\neq p_1,\dots, p_n} {1 \over p}\label{eq5p1}
\end{align}
We have already noted above that the appropriate regularization scheme for the case of $n=2$ is (\ref{twoexclusions}) with $x= {1\over 3}$. In this section, we analyze the case of general $n$. In particular, we prove that the replacement
\begin{align}
 \sum'_{p\neq p_1,\dots, p_n} {1 \over p}\longrightarrow Q_1(p_1,\dots,p_n) = -\sum_{i=1}^n {1 \over p_i} - {\pi \over (n+1) \tau_2}\sum_{i=1}^n(p_i - \bar p_i )
\end{align}
yields an expression which is modular covariant, and which matches with the result obtained using a different regularization procedure. Once we have proven this, it is straightforward (though tedious) to obtain $(n+1)$-point holomorphic subgraph reduction formulae for graphs of any order $n+1$.

\subsection{Definition of $Q_1(p_1,p_2)$}
We begin by recalling the context in which $Q_1(p_1,p_2)$ appears in the three-point case just studied. The reason these divergent sums arise in an otherwise finite calculation is that we have decomposed the original, absolutely convergent sum (\ref{sumS}) into a number of divergent or conditionally convergent sums. This led us to define $Q_1(p_1,p_2)$ and $Q_2(p_1,p_2)$, which were taken to be given by (\ref{twoexclusions}) with one free parameter $x$. 

If we were only interested in regularizing (\ref{eq5p1}), there would be no preferred value of $x$. However, the fact that the terms $Q_1(p_1, p_2)$ and $Q_2(p_1, p_2)$ arise from the decomposition of (\ref{sumS}) allows us to put physical constraints on $x$. In particular, we note that (\ref{sumS}) is modular covariant, and that this is a property that we would like our regularization procedure to preserve. We now note that the ${\pi \over \tau_2} p_i$ terms in $Q_1(p_1, p_2)$, as well as the ${\pi \over \tau_2}$ term in $Q_2(p_1, p_2)$, are of a different modular weight than the other terms -- in particular, if the original modular graph form carried modular weight ${(w,\bar w)}$, these abnormal terms are of modular weight ${(w-1,\bar w+1)}$. Thus the correct choice of $x$ for the current purposes is the one for which the terms of abnormal modular weight ${(w-1,\bar w+1)}$ cancel when combined in (\ref{sumS}). Imposing this constraint fixes $x={1/3}$.
 
\subsection{Definition of $Q_1(p_1,p_2,p_3)$}
Before examining the general case, we offer one more explicit example. In particular, we use the same analysis as above to determine $Q_1(p_1,p_2,p_3)$. The natural starting point is the consideration of four-point holomorphic subgraphs which arise in tetrahedral modular graph forms. Such a four-point holomorphic subgraph is shown below, together with our momentum orientation conventions, 
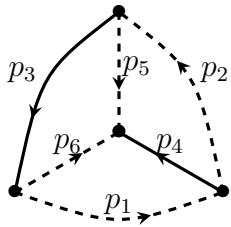
\begin{align}
\label{fig15}
\tikzpicture[scale=0.8]
\scope[xshift=-5cm,yshift=-0.4cm]
\draw[directed, very thick, dashed]  (7.73,-1) ..controls (7.3,1) .. (6,2);
\draw[directed, very thick] (6,2) ..controls (4.7,1) ..(4.27,-1);
\draw[directed, very thick,dashed] (4.27,-1)  ..controls (6,-1.6) .. (7.73,-1);
\draw[ directed, very thick,dashed] (6,2) node{$\bullet$} --  (6,0) node{$\bullet$};
\draw[ directed, very thick,dashed] (4.27,-1) node{$\bullet$} --  (6,0) node{$\bullet$};
\draw[ directed, very thick] (7.73,-1) node{$\bullet$} --  (6,0) node{$\bullet$};
\draw (7.6,1) node{$p_2$};
\draw (4.4,1) node{$p_3$};
\draw (6,-1.2) node{$p_1$};
\draw (6.3,1.1) node{$p_5$};
\draw (5.15,-0.2) node{$p_6$};
\draw (6.85,-0.2) node{$p_4$};
\endscope
\endtikzpicture
\no
\end{align}
Instead of considering the general form of tetrahedral graphs with four-point holomorphic subgraphs, we will consider only a particular weight ${(3,-3)}$ example,
\begin{align}
 \cC\left[\begin{matrix} 2 \\ 0\end{matrix} \middle| \begin{matrix} 2 \\ 0\end{matrix} \middle| \begin{matrix} 1 \\ 1\end{matrix} \middle| \begin{matrix} 1 \\ 1\end{matrix}  \middle| \begin{matrix} 1 \\ 0\end{matrix} \bigg | \begin{matrix} 1 \\ 0\end{matrix}\right]= \sum'_{p_i \in \Lambda} \delta_{p_1, p_2+p_4} \delta_{p_2, p_3 + p_5} \delta_{p_3, p_1 + p_6} \left({\tau_2 \over \pi} \right)^{5}{1 \over p_3 \bar p_3 p_4 \bar p_4}\,{1 \over p_1^2 p_2^{2} p_5 p_6}
\end{align}
Similar to the technique used in the trihedral case, we begin by expressing all of the momenta in terms of $p_1$ and the external momenta $p_3, p_4$ as follows 
\begin{align}
p_2 &= p_1 - p_4 & p_1 &\neq p_4
\no\\
p_6 &= p_3 - p_1 & p_1 &\neq p_3
\no\\
p_5 &= p_1 - p_3 - p_4 & p_1&\neq p_3 + p_4
\end{align}
The sums of the form $Q_1(p_1,p_2,p_3)$ then arise when we decompose 
\begin{align}
\label{tetrahedral1}
&\sum'_{\substack{p_1 \neq p_3, p_4 \\p_{1}\neq p_3 + p_4 }} {1 \over p_1^2 (p_3 - p_1)(p_1 - p_4)^{2}(p_1 - p_3 - p_4)}
\no\\
 &\vphantom{.}=\hspace{-0.5em}\sum'_{\substack{p_1 \neq p_3, p_4 \\p_{1}\neq p_3 + p_4 }}\hspace{-0.5em}  \left[{1 \over p_3 p_4^2 (p_4 - p_3)}\,{1 \over (p_1 - p_4)^{2}} +{1 \over p_4 p_3^2 (p_3 - p_4)^{2}}\,{1 \over p_1 - p_3}-{1 \over p_3^{2} p_4 (p_3 + p_4)^2}\,{1 \over p_1 - p_3 - p_4} \right.
\no\\
&\hphantom{\sum'_{\substack{p_1 \neq p_3, p_4 \\p_{1}\neq p_3 + p_4 }}  \Big[}\left. +\frac{2p_{3}^{2}-4p_{3}p_{4}+p_{4}^{2}}{p_{3}^{2}p_{4}^{3}(p_{3}-p_{4})^{2}}\,\frac{1}{p_{1}-p_{4}} - {2p_3^2 + 4 p_3 p_4 + p_4^2 \over p_3^2 p_4^3 (p_3 + p_4)^2}\,{1 \over p_1} - {1 \over p_3 p_4^{2} (p_3 + p_4)}\,{1 \over p_1^2}   \right]
\no\\
\end{align}
As before, we evaluate the sums over $p_1$ in terms of $Q_1$ and $Q_2$, with the natural ansätze
\begin{align}
\label{tetraansatze}
Q_1(p_1, p_2, p_3) &= -{1 \over p_1} - {1 \over p_2}-{1 \over p_3} -x{\pi \over \tau_2}(p_1 -\bar p_1 + p_2 -\bar p_2+ p_3 -\bar p_3 )
\no\\
Q_2(p_1, p_2, p_3) &=   - {1 \over p_1^2} - {1 \over p_2^2}-{1 \over p_3^2}+\mathcal{\hat G}_2 + {\pi \over \tau_2}
\end{align}
From now on, we will keep only the $x{\pi \over \tau_2} p_i$ terms from $Q_1$ and the ${\pi \over \tau_2}$ term from $Q_2$ which carry incorrect modular weight $(2,-2)$. The appropriate regularization is again the one such that these terms cancel. 

In total, one finds the following contribution to (\ref{tetrahedral1}) from such terms, 
\begin{align}
  \frac{2}{p_{4}^{2}(p_{3}^{2}-p_{4}^{2})}\,\frac{\pi}{\tau_{2}}(4x-1)
\end{align}
from which we conclude that $x=1/4$ gives the appropriate regularization.

Importantly, the same conclusion holds no matter which tetrahedral modular graph form we are performing the four-point holomorphic subgraph reduction on. For example, we could instead have begun with
\begin{align}
 \cC\left[\begin{matrix} 2 \\ 0\end{matrix} \bigg| \begin{matrix} 2 \\ 0\end{matrix} \bigg | \begin{matrix} 1 \\ 1\end{matrix} \bigg | \begin{matrix} 1 \\ 1\end{matrix}  \bigg | \begin{matrix} 2 \\ 0\end{matrix} \bigg | \begin{matrix} 2 \\ 0\end{matrix}\right] = \sum'_{p_i \in \Lambda} \delta_{p_1, p_2+p_4} \delta_{p_2, p_3 + p_5} \delta_{p_3, p_1 + p_6} \left({\tau_2 \over \pi} \right)^{6}{1 \over p_3 \bar p_3 p_4 \bar p_4}{1 \over p_1^2 p_2^2 p_5^2 p_6^2}
\end{align}
in which case the sum of interest is 
\begin{align}
\sum'_{\substack{p_1 \neq p_3, p_4, \\ p_3 + p_4 }} {1 \over p_1^2 (p_1 - p_3)^2(p_1 - p_4)^2(p_1 - p_3 - p_4)^2}
\end{align}
Exactly analogous steps confirm that in this case as well, $x=1/4$ is the correct choice. Proving that this is a general feature is the goal of the next subsection.

\subsection{Definition of general $Q_1(p_1,\dots, p_n)$}
\label{sec:genreg}
The general strategy is now clear. The appearance of $Q_1(p_1,\dots, p_n)$ in holomorphic subgraph reduction of modular graph forms always comes from the decomposition of sums of the form 
\begin{align}
\label{arbitraryexps}
\sum'_{p \neq p_{1}, \dots, p_{n}}{1 \over p^{a_0} (p - p_1)^{a_1} \dots (p-p_n)^{a_n}}
\end{align}
for some external momenta $p_i$ and corresponding exponents $a_i$, $i=1,\dots,n$. We will assume that all of the $p_i$ are distinct; if this is not the case, we can just increase the corresponding exponents. We will also exclude the case of $n=1$, $a_{0}=a_{1}=1$, since in that case the sum \eqref{arbitraryexps} is not absolutely convergent, and does not appear in any physical calculations. 

It suffices to specialize to the case\footnote{We may get a sum (\ref{arbitraryexps}) with arbitrary $a_i$, $i=1,\dots,n$ from (\ref{2111exps}) by differentiating with respect to the external momenta $p_i$. The validity of this interchange of derivatives and sums follows by uniform convergence. }
\begin{align}
\label{2111exps}
\sum'_{p \neq p_{1}, \dots, p_{n}} {1 \over p^{a_0} (p - p_1) \dots (p-p_n)}
\end{align}
For any $a_{0},n\geq1$ we may now use a partial fraction decomposition to re-express\footnote{This can be proven by induction over $n$ either directly, or alternatively by using the relations
\begin{align}
  \prod_{i=1}^{n}\frac{1}{p-p_{i}}=\sum_{i=1}^{n}\frac{1}{p-p_{i}}\prod_{\substack{j=1\\j\neq i}}^{n}\frac{1}{p_i- p_j}\label{eq:15}
\end{align}
and 
\begin{align}
\label{usefulidentities}
\sum_{i=1}^n {1 \over p_i^{a_0}} \prod_{\substack{j=1\\j \neq i}}^n{1 \over p_i - p_j} = (-1)^{n+1} h_{a_0-1}(p_1,\dots,p_n) \prod_{i=1}^n{1\over p_i^{a_0}}
\end{align}
which may themselves be verified by inductive arguments.}
\begin{align}
\sum'_{p \neq p_{1}, \dots, p_{n}}\left[{1 \over p^{a_0} } \prod_{i=1}^n {1 \over p-p_i}\right] =\sum'_{p \neq p_{1}, \dots, p_{n}}\left[ \sum_{i=1}^n \left( {1 \over p_i^{a_0} (p-p_i)}\prod_{\substack{j=1\\j \neq i}}^n {1 \over p_i - p_j}\right) +(-1 )^n \sum_{\ell=1}^{a_0}{ h_{\ell-1}(p_1,\dots,p_n)\over p^{a_0 - \ell+1}\prod_{i=1}^n{p_i^\ell}}\right] \label{eq:14}
\end{align}
The  $h_{k}(p_1,\dots,p_n)$ are symmetric polynomials in $p_1, \dots, p_n$ of homogeneous order $(n-1) k$, defined by the following expression,
\begin{align}
h_k(p_1,\dots, p_n) = \sum_{\substack{a_1,\dots, a_n = 0 \\ a = (n-1) k}}^k \prod_{i=1}^n p_i^{a_i}
\end{align}
with $a =  a_1 +\dots + a_n$.\footnote{Note that unlike in the previous sections, $a_0$ is now being used to refer to a single exponent, as opposed to a sum over them. }

One way to carry out the summation over $p$ in \eqref{eq:14} is to choose a summation prescription for which the sum over each individual term in the summand converges, and then to distribute the sum over the individual terms. In particular, we may work with the Eisenstein summation prescription, denoted by $\esum$ and defined in (\ref{eq:27}) of Appendix \ref{sec:eisenst-regul-simple}, and then distribute the sums in \eqref{eq:14}, yielding
\begin{align}
  \sum_{i=1}^n \left(\hspace{0.5em}\esum'_{p \neq p_{1}, \dots, p_{n}}\frac{1}{p-p_{i}} {1 \over p_i^{a_0}}\prod_{\substack{j=1\\j \neq i}}^n {1 \over p_i - p_j}\right) +(-1 )^n \sum_{\ell=1}^{a_0}\hspace{0.5em}\esum'_{p \neq p_{1}, \dots, p_{n}}\frac{1}{p^{a_{0}-\ell+1}}{ h_{\ell-1}(p_1,\dots,p_n)\over \prod_{i=1}^n{p_i^\ell}} 
\end{align}
The contributions to this with one power of $p$ in the denominator may be evaluted using the identities \eqref{eq:20},\eqref{eq:22}, and \eqref{usefulidentities} to give
\begin{align}
  \sum_{i=1}^{n}\left[\left(\frac{2}{p_{i}}+\sum_{\substack{j=1\\i\neq j}}^{n}\left(\frac{1}{p_{i}-p_{j}}+\frac{1}{p_{j}}\right)+\frac{\pi}{\tau_{2}}(p_{i}-\bar p_{i})\right)\frac{1}{p_{i}^{a_{0}}}\prod_{\substack{j=1\\i\neq j}}^{n}\frac{1}{p_{i}-p_{j}}\right]\label{eq:23}
\end{align}
Since the derivation of this result only involved the evaluation of convergent sums, no ambiguity was introduced and hence \eqref{eq:23} is the same function of $\tau$ as \eqref{2111exps}.

A second method to carry out the summation over $p$ in \eqref{eq:14} is to just distribute the sum over each individual term in the summand regardless of convergence, and then to regularize each of the individual sums in an appropriate way. This is the situation in which the $Q_{1}(p_1,\dots,p_n)$ arise. We would now like to choose a definition for  $Q_{1}(p_1,\dots,p_n)$ such that we reproduce the result \eqref{eq:23}. The correct replacements for this matching are
\begin{align}
  \sum_{p\neq p_{1},\dots,p_n}' \frac{1}{p}&\longrightarrow Q_{1}(p_{1},\dots,p_n)\label{eq:29}\\
  \sum_{p\neq p_{1},\dots,p_n}' \frac{1}{p_{i}-p}&\longrightarrow Q_{1}(p_{i},\underbrace{p_{i}-p_{1},\dots,p_{i}-p_{n}}_{\text{omit $p_{i}-p_{i}$}})\label{eq:31}
\end{align}
with $  Q_1(p_1, \dots, p_n) $ given by
\begin{align}
  Q_1(p_1, \dots, p_n) &= -\sum_{\ell=1}^n{1 \over p_\ell} -{\pi \over(n+1) \tau_2}\sum_{\ell=1}^n(p_\ell -\bar p_\ell)\label{eq:24}
\end{align}
That this leads indeed to \eqref{eq:23} can be straightforwardly verified and hence using \eqref{eq:24} in holomorphic subgraph reduction will produce correct identities.

In contrast to direct Eisenstein summation of the original sum, where shifted sums had to be treated separately and generated additional terms, the replacements \eqref{eq:29} and \eqref{eq:31} are very intuitive and simple to implement. In particular, shifted and unshifted sums can be regularized using the same expression. Therefore, for practical calculations, regularizing divergent sums using the $Q_{1}(p_1, \dots, p_n)$ is preferable.

This form of $Q_1(p_1, \dots, p_n)$ can also be obtained via Eisenstein summation of a certain linear combination of shifted sums, as outlined in Appendix \ref{sec:q_1shiftedsums}. 

Note that although the sum
\begin{align}
  \sum_{p\neq p_{1},\dots,p_n}' \frac{1}{p^{2}}
\end{align}
is only conditionally convergent, it does not have the same shift-dependence under Eisenstein summation as $\sum 1/p$, and hence the definition of $Q_{2}(p_{1},\dots,p_{n})$ does not suffer from the same ambiguities as $Q_{1}(p_{1},\dots,p_{n})$. For details see Appendix \ref{sec:eisenst-regul-simple}.

We now show that the regularization identified above,
\begin{align}
  Q_1(p_1, \dots, p_n) &= -\sum_{\ell=1}^n{1 \over p_\ell} -{\pi \over(n+1) \tau_2}\sum_{\ell=1}^n(p_\ell -\bar p_\ell)
\no\\
  Q_2(p_1,\dots, p_n) &= - \sum_{\ell=1}^n {1 \over p_\ell^2}+ {\rm \hat \cG}_2 + {\pi \over \tau_2}\no\\
  Q_{k}(p_{1},\dots,p_{n})&={\rm \cG}_{k}- \sum_{\ell=1}^n {1 \over p_\ell^k}\quad k\geq3\label{eq:25}
\end{align}
leads to a modular covariant final result for (\ref{2111exps}). Since we are interested only in the terms of abnormal modular weight, we may discard all terms in the sum over $\ell$ for which $a_0 - \ell + 1 >2$. For $a_{0}\geq2$ we insert \eqref{eq:25} into \eqref{eq:14} and keep only the terms $-{\pi \over (n+1)\tau_2}\sum_{\ell=1}^np_\ell$ in $Q_{1}$ and $\frac{\pi}{\tau_{2}}$ in $Q_{2}$, to obtain\footnote{For $a_{0}=1, n\geq2$, one finds $F(p_{1},\dots,p_{n};1)=0$ and the second term in \eqref{setzero} is absent. }
\begin{align}
\label{setzero}
{\tau_2 \over \pi} \sum'_{p \neq p_{1}, \dots, p_{n}} {1 \over p^{a_0} (p - p_1) \dots (p-p_n)} \Big|_\text{abnorm} = \frac{1}{n+1} F(p_1, \dots, p_n ; a_0) +(-1)^n h_{a_0-2}(p_1,\dots,p_n) \prod_{i=1}^n{1 \over p_i^{a_0-1}}
\end{align}
where we have defined 
\begin{align}
F(p_1, \dots, p_n ; a_0)&=\sum_{i=1}^n {1 \over p_i^{a_0}  \prod_{\substack{j=1\\j \neq i}}^n(p_i - p_j) } \left[p_i - \sum_{\substack{k=1 }}^n (p_k - p_i) \right]\no\\
                        &\hphantom{=}\hspace{1 in}+(-1)^{n+1} h_{a_0-1}(p_1,\dots, p_n)\left( \prod_{i=1}^n { 1\over p_i^{a_0}} \right)\sum_{j=1}^n p_j\label{eq:17}
\end{align}
The first line of (\ref{eq:17}) can be rewritten as
\begin{align}
  &\sum_{i=1}^n {1 \over p_i^{a_0}  \prod_{\substack{j=1\\j \neq i}}^n(p_i - p_j) } \left[p_i - \sum_{\substack{k=1}}^n (p_k - p_i) \right]\no\\
 &\vphantom{.} \hspace{0.5 in}= (n+1) \sum_{i=1}^n {1 \over p_i^{a_0-1}} \prod_{\substack{j=1\\j \neq i}}^n{1 \over p_i - p_j}  - \sum_{i=1}^n {1 \over p_i^{a_0}}\prod_{\substack{j=1\\j \neq i}}^n{1 \over p_i - p_j} \sum_{\substack{k=1}}^n  p_k\label{eq:16}
\end{align}
Applying \eqref{usefulidentities} to both terms of \eqref{eq:16} and plugging this back into \eqref{eq:17} yields
\begin{align}
\label{Claim}
F(p_1, \dots, p_n ; a_0) = (-1)^{n+1} (n+1) h_{a_0-2}(p_1,\dots,p_n)  \prod_{i=1}^n {1 \over p_i^{a_0-1}}
\end{align}
Using this in \eqref{setzero} then gives
\begin{align}
  {\tau_2 \over \pi} \sum'_{p \neq p_{1}, \dots, p_{n}} {1 \over p^{a_0} (p - p_1) \dots (p-p_n)} \Big|_\text{abnorm} = 0
\end{align}
which confirms that for the $Q_{k}$ defined in \eqref{eq:25} the terms of abnormal modular weight do indeed cancel out, and the final result of the holomorphic subgraph reduction procedure is modular covariant.

\section{Summary}
In this work, we have extended the results of \cite{DHoker:2016mwo} to obtain holomorphic subgraph reduction formulae for trihedral modular graph forms. The two-point holomorphic subgraph reduction formula was given in (\ref{trivalent2pt}), and is a simple generalization of the two-point formula for dihedral modular graph forms. The three-point holomorphic subgraph reduction formula is considerably more involved, and was given in (\ref{eq:9}). It involves a sum over five pieces, which were given in  (\ref{eq:5})-(\ref{cL4def}) and (\ref{finalL5}). While this result seems rather involved, it is easy to implement digitally, and has already yielded physically useful results in the context of heterotic string amplitudes \cite{upcoming2}. 

The method by which we obtained these results involved the decomposition of absolutely convergent sums into a number of divergent or conditionally convergent sums, for which the appropriate, modular covariant regularization scheme \eqref{eq:25} was identified. With this in hand, it is straightforward to perform holomorphic subgraph reduction on higher-point modular graph forms with arbitrary holomorphic subgraphs. The extension of the explicit formulae to the general case is left to the ambitious reader.

\addtocontents{toc}{\protect\setcounter{tocdepth}{0}}
\section*{Acknowledgments}
\addtocontents{toc}{\protect\setcounter{tocdepth}{2}}
We thank Axel Kleinschmidt, Oliver Schlotterer, Eric D'Hoker, and Bill Duke for enlightening discussions. J.G. also thanks Axel Kleinschmidt and Oliver Schlotterer for collaboration on related topics that initiated the present work. J.G.\ is supported by the International Max Planck Research School for Mathematical and Physical Aspects of Gravitation, Cosmology and Quantum Field Theory. J.K. would like to thank the Yukawa Institute for Theoretical Physics and the Simons Center for Geometry and Physics for their hospitality during the completion of this work, and the Mani L. Bhaumik Institute for Theoretical Physics for generous support.

\appendix

\section{Trihedral holomorphic subgraph reduction without $Q_{i}(p_{1},\narrowdots,p_{n})$}\label{sec:two-exampl-carefully}
In this section we outline derivations of the decompositions \eqref{eq:12} and \eqref{eq:13} which do not involve divergent sums which must be regularized. This will serve as a check for the consistency of our regularization procedure. Note that derivations of this sort must be found on a case-by-case basis, and do not admit a nice systematization like that studied in the main text.

First, consider the sum
\begin{align}
  \left(\frac{\pi}{\tau_{2}}\right)^{4}\cformtri{1&1\\1&0}{1&2\\1&0}{1\\0}=\sum'_{\substack{p_{1},p_{2},p_{3}\\p_{1}+p_{2}\neq0\\p_{1}+p_{3}\neq0}}\frac{1}{|p_{2}|^{2}|p_{3}|^{2}}\frac{1}{p_{1}(p_{1}+p_{2})(p_{1}+p_{3})^{2}}
\end{align}
Using the decomposition
\begin{align}
  \frac{1}{p_{1}(p_{1}+p_{2})(p_{1}+p_{3})^{2}}=\left(\frac{1}{p_{1}p_{2}}-\frac{1}{p_{2}(p_{1}+p_{2})}\right)\frac{1}{(p_{1}+p_{3})^{2}}
\end{align}
and including and subtracting the terms with $p_{1}+p_{2}=0$ in the first sum and the terms with $p_{1}=0$ in the second sum, this can be rewritten as
\begin{align}
  \left(\frac{\pi}{\tau_{2}}\right)^{4}\cformtri{1&1\\1&0}{1&2\\1&0}{1\\0}=&\sum'_{\substack{p_{1},p_{2},p_{3}\\p_{1}+p_{3}\neq0}}\frac{1}{|p_{2}|^{2}|p_{3}|^{3}}\frac{1}{p_{1}p_{2}(p_{1}+p_{3})^{2}}+\sum'_{\substack{p_{2},p_{3}\\p_{2}\neq p_{3}}}\frac{1}{|p_{2}|^{2}|p_{3}|^{2}}\frac{1}{p_{2}^{2}(p_{2}-p_{3})^{2}}\no\\
  &-\sum_{\substack{p_{1},p_{2},p_{3}\\p_{2},p_{3}\neq0\\p_{1}+p_{2}\neq0\\p_{1}+p_{3}\neq0}}\frac{1}{|p_{2}|^{2}|p_{3}|^{2}}\frac{1}{p_{2}(p_{1}+p_{2})(p_{1}+p_{3})^{2}}+\sum'_{p_{2},p_{3}\neq0}\frac{1}{|p_{2}|^{2}|p_{3}|^{2}}\frac{1}{p_{2}^{2}p_{3}^{2}}\label{eq:6}
\end{align}
In the first term, the sum over $p_{2}$ factorizes and vanishes by antisymmetry. The second term is dihedral and the last term factorizes completely. The third term can be shown to be a dihedral modular graph form by relabeling $p_{1}\rightarrow p_{1}-p_{3}$ and then $p_{3}\rightarrow -p_{3}$,
\begin{align}
  \sum_{\substack{p_{1},p_{2},p_{3}\\p_{2},p_{3}\neq0\\p_{1}+p_{2}\neq0\\p_{1}+p_{3}\neq0}}\frac{1}{|p_{2}|^{2}|p_{3}|^{2}}\frac{1}{p_{2}(p_{1}+p_{2})(p_{1}+p_{3})^{2}}&=\sum'_{\substack{p_{1},p_{2},p_{3}\\p_{1}+p_{2}+p_{3}\neq0}}\frac{1}{|p_{2}|^{2}|p_{3}|^{2}}\frac{1}{p_{1}^{2}p_{2}(p_{1}+p_{2}+p_{3})}\no\\
  &=-\left(\frac{\pi}{\tau_{2}}\right)^{4}\cform{1&2&1&2\\0&0&1&1}
\end{align}
Collecting all terms, we obtain the decomposition
\begin{align}
  \cformtri{1&1\\1&0}{1&2\\1&0}{1\\0}=\cform{3&0\\1&0}^{2}+\cform{3&1&2\\1&1&0}+\cform{1&2&1&2\\0&0&1&1}
\end{align}
Using dihedral holomorphic subgraph reduction on the last term confirms the earlier result \eqref{eq:12}. Note that in this derivation, all the sums appearing in every step were absolutely convergent and no regularization was needed. 

As a second example, consider
\begin{align}
  \left(\frac{\pi}{\tau_{2}}\right)^{4}\cformtri{2\\0}{1&1\\0&1}{1&1\\0&1}=\sum_{\substack{p_{1},p_{2},p_{3}\\p_{1}+p_{2}\neq0\\p_{1}+p_{3}\neq0}}' \frac{1}{|p_{2}|^{2}|p_{3}|^{2}}\frac{1}{p_{1}^{2}(p_{1}+p_{2})(p_{1}+p_{3})}
\end{align}
We begin by reorganizing $p_{1}$ as $p_{4}\equiv p_{1}+p_{2}$, resulting in
\begin{align}
  \sum_{\substack{p_{2},p_{3},p_{4}\\p_{4}-p_{2}\neq0\\p_{4}-p_{2}+p_{3}\neq0}}'\frac{1}{|p_{2}|^{2}|p_{3}|^{2}}\frac{1}{p_{4}(p_{4}-p_{2})^{2}(p_{4}-p_{2}+p_{3})}
\end{align}
Using partial fraction decomposition, this can be transformed into
\begin{align}
  \sum_{\substack{p_{2},p_{3},p_{4}\\p_{4}-p_{2}\neq0\\p_{4}-p_{2}+p_{3}\neq0}}'\frac{1}{|p_{2}|^{2}|p_{3}|^{2}}\left(\frac{1}{p_{2}(p_{4}-p_{2})^{2}(p_{4}-p_{2}+p_{3})}-\frac{1}{p_{2}p_{4}(p_{4}-p_{2})(p_{4}-p_{2}+p_{3})}\right)
\end{align}
We now return to $p_{1}=p_{4}-p_{2}$ in the first term and decompose the second term once more, leading to
\begin{align}
  &\sum_{\substack{p_{1},p_{2},p_{3}\\p_{1}+p_{2}\neq0\\p_{1}+p_{3}\neq0}}' \frac{1}{|p_{2}|^{2}|p_{3}|^{2}}\frac{1}{p_{1}^{2}p_{2}(p_{1}+p_{3})}\notag\\
  &\vphantom{.} \hspace{0.5 in}-\sum_{\substack{p_{2},p_{3},p_{4}\\p_{4}-p_{2}\neq0\\p_{4}-p_{2}+p_{3}\neq0}}'\frac{1}{|p_{2}|^{3}|p_{3}|^{2}}\left(\frac{1}{p_{2}p_{4}^{2}(p_{4}-p_{2}+p_{3})}+\frac{1}{p_{3}p_{4}^{2}(p_{4}-p_{2})}-\frac{1}{p_{3}p_{4}
  ^{2}(p_{4}-p_{2}+p_{3})}\right)\label{eq:19}
\end{align}
Evaluating the first term leads to
\begin{align}
  \sum_{\substack{p_{1},p_{2},p_{3}\\p_{1}+p_{2}\neq0\\p_{1}+p_{3}\neq0}}' \frac{1}{|p_{2}|^{2}|p_{3}|^{2}}\frac{1}{p_{1}^{2}p_{2}(p_{1}+p_{3})}&=\sum_{\substack{p_{1},p_{2},p_{3}\\p_{1}+p_{3}\neq0}}' \frac{1}{|p_{2}|^{2}|p_{3}|^{2}}\frac{1}{p_{1}^{2}p_{2}(p_{1}+p_{3})}+\sum_{\substack{p_{1},p_{3}\\p_{1}+p_{3}\neq0}}' \frac{1}{|p_{1}|^{2}|p_{3}|^{2}}\frac{1}{p_{1}^{3}(p_{1}+p_{3})}
\end{align}
The first term in the above vanishes due to antisymmetry in $p_{2}$, while the second one is dihedral, and hence
\begin{align}
  \sum_{\substack{p_{1},p_{2},p_{3}\\p_{1}+p_{2}\neq0\\p_{1}+p_{3}\neq0}}' \frac{1}{|p_{2}|^{2}|p_{3}|^{2}}\frac{1}{p_{1}^{2}p_{2}(p_{1}+p_{3})}=-\left(\frac{\pi}{\tau_{2}}\right)^{4}\cform{4&1&1\\1&1&0}
\end{align}
Similarly, we evaluate the second term in \eqref{eq:19}, yielding
\begin{align}
  &\sum_{\substack{p_{2},p_{3},p_{4}\\p_{4}-p_{2}\neq0\\p_{4}-p_{2}+p_{3}\neq0}}'\frac{1}{|p_{2}|^{3}|p_{3}|^{2}}\left(\frac{1}{p_{2}p_{4}^{2}(p_{4}-p_{2}+p_{3})}+\frac{1}{p_{3}p_{4}^{2}(p_{4}-p_{2})}-\frac{1}{p_{3}p_{4}
  ^{2}(p_{4}-p_{2}+p_{3})}\right)\notag\\
  &\vphantom{.} \hspace{0.5 in}=\left(\frac{\pi}{\tau_{2}}\right)^{4}\left\{2\cform{2&1&2&1\\1&1&0&0}+\cform{1&3&2\\1&1&0}+\cform{3&0\\1&0}^{2}\right\}
\end{align}
Putting everything together yields the final expression
\begin{align}
  \cformtri{2\\0}{1&1\\0&1}{1&1\\0&1}=-2\cform{2&1&2&1\\1&1&0&0}-\cform{4&1&1\\1&1&0}-\cform{1&3&2\\1&1&0}-\cform{3&0\\1&0}^{2}
\end{align}
which, upon dihedral holomorphic subgraph reduction of the first term and usage of some further dihedral identities, yields \eqref{eq:13}.
 
 \newpage
\section{Eisenstein summation}
\label{tworegs}

\subsection{Eisenstein summation of simple sums}
\label{sec:eisenst-regul-simple}
In this Appendix, we apply the Eisenstein summation prescription to sums which are needed to evaluate the expression \eqref{eq:14}. The Eisenstein summation prescription is defined as follows,
\begin{align}
  \esum_{p\neq r+s\tau} f(p)\equiv&\lim_{N\rightarrow\infty}\sum_{\substack{n=-N\\n\neq s}}^{N}\left(\lim_{M\rightarrow\infty}\sum_{m=-M}^{M}f(m+n\tau)\right)+\lim_{M\rightarrow\infty}\sum_{\substack{m=-M\\m\neq r}}^{M}f(m+s \tau)\label{eq:27}
\end{align}
where $f$ is assumed to have a pole at $r+s\tau\in\Lambda$ and takes on finite values at all other lattice points. If a finite number of additional points are excluded from the sum, they have to be subtracted from the right-hand side.

We first consider the case $f(p)=1/{p}$ with the points $0$ and $P\equiv\left\{p_{i}=m_{i}+n_{i}\tau \,\big| \,i=1,\dots,n\right\}$ being excluded from the sum. Then the sum over $1/{m}$ vanishes by antisymmetry. For the sum over $\frac{1}{m+n\tau}$, we use the trigonometric identity
\begin{align}
    \lim_{M\rightarrow\infty}\sum_{m=-M}^{M}\frac{1}{m+n\tau}=-i\pi\frac{1+q^{n}}{1-q^{n}}\label{eq:28}
\end{align}
The sum of this over $n$ also vanishes by antisymmetry. Hence, the only remaining term is due to the excluded points in $P$ and we obtain
\begin{align}
    \esum'_{p\notin P}\frac{1}{p}=-\sum_{p\in P}\frac{1}{p}\label{eq:20}
\end{align}
In a similar fashion, we may now consider the case in which $f(p)=1/(p_{i}-p)$ and the points in $P\cup\{0\}$ are excluded. Again using vanishing of the sum over $1/m$, as well as \eqref{eq:28}, we have
\begin{align}
  \esum_{p\notin P}'\frac{1}{p_{i}-p}=-\frac{1}{p_{i}}-\sum_{\substack{p\in P\\p\neq p_{i}}}\frac{1}{p_{i}-p}-i\pi\lim_{N\rightarrow\infty}\sum_{\substack{n=-N+n_{i}\\n\neq0}}^{N+n_{i}}\frac{1+q^{n}}{1-q^{n}}
\end{align}
Now the final sum does not vanish due to the asymmetric summation range. Indeed, noting that
\begin{align}
    \frac{1+q^{n}}{1-q^{n}}\longrightarrow\pm1\quad(n\rightarrow\pm\infty)\label{eq:30}
\end{align}
we obtain the result
\begin{align}
    \esum_{p\notin P}'\frac{1}{p_{i}-p}&=-\frac{1}{p_{i}}-\sum_{\substack{p\in P\\p\neq p_{i}}}\frac{1}{p_{i}-p}-\frac{\pi}{\tau_{2}}(p_{i}-\bar p_{i})\label{eq:22}
\end{align}

Note that this illustrates an important point: upon Eisenstein summation the sum $\sum_{p}' \frac{1}{p}$ is not invariant under shifts of the summation variable $p\rightarrow p-p_{i}$. This effect does not occur for sums of the form $\sum_{p}' \frac{1}{p^{k}}$ with $k\geq 3$ since they are absolutely convergent. For the case $k=2$, the sum is conditionally convergent. But because
\begin{align}
  \lim_{M\rightarrow\infty}\sum_{m=-M}^{M}\frac{1}{(m+n\tau)^{2}}=-4\pi^{2}\frac{q^{n}}{(1-q^{n})^{2}}\longrightarrow0\quad(n\rightarrow\pm\infty)
\end{align}
the regularization in this case is found to not be shift dependent, leaving us with the result
\begin{align}
    Q_{2}(p_{1},\dots,p_{n})\equiv\esum'_{p\neq p_{1},\dots,p_{n}}\frac{1}{p^{2}}=-\sum_{i=1}^{n}\frac{1}{p_{i}^{2}}+\mathcal{\hat G}_{2}+\frac{\pi}{\tau_{2}}
\end{align}

\subsection{$Q_1(p_1,\dots,p_n)$ from shifted sums}
\label{sec:q_1shiftedsums}
We now obtain the expression for $Q_{1}(p_1,\dots,p_n)$ given in \eqref{eq:25} of the main text by Eisenstein summing a certain linear combination of shifted sums. We start by defining
\begin{align}
Q_1 (p_1, \dots, p_n) \equiv \half \left[\hspace{0.5em}\esum'_{p \neq p_1, \dots, p_n} {1 \over p} + {1 \over n+1} \sum_{i=1}^n \esum_{\substack{p\neq p_{i}\\p\notin P_{i}}} {1 \over p_i - p} + {1 \over n+1} \esum_{\substack{p\neq p_{0}\\p\notin P_{0}}} {1 \over p_0 - p }\right] \label{eq:18}
\end{align}
where we have defined $p_0 \equiv \sum_{i = 1}^n p_i$, as well as the set $P_{k}\equiv\{p_{k}-p_{j}|j=1,\dots,n\}$. Using the identities \eqref{eq:20} and \eqref{eq:22} from the previous section, we can evaluate \eqref{eq:18}, yielding
\begin{align}
  Q_1(p_1, \dots, p_n) &= -\frac{1}{2}\left[\sum_{i=1}^{n}\frac{1}{p_{i}}+\frac{1}{n+1}\sum_{i=1}^{n}\sum_{j=1}^{n}\frac{1}{p_{j}}+\frac{1}{n+1}\sum_{j=1}^{n}\frac{1}{p_{j}}\right.\no\\
  &\hspace{0.5 in}\hphantom{=-\frac{1}{2}\Big[}\left.+\frac{\pi}{(n+1)\tau_{2}}\left(\sum_{i=1}^{n}(p_{i}-\bar p_{i})+p_{0}-\bar p_{0}\right)\right]
\no\\
&= - \sum_{i=1}^n {1 \over p_i}  - {\pi \over (n + 1) \tau_2} \sum_{i=1}^n \left(p_i - \bar p_i \right)
\end{align}


\newpage
\begin{thebibliography}{99}

\bibitem{DHoker:2016mwo} 
  E.~D'Hoker and M.~B.~Green,
  ``Identities between Modular Graph Forms,''
  J.\ Number Theor.\  {\bf 189}, 25 (2018)
  [arXiv:1603.00839 [hep-th]].

\bibitem{upcoming2} 
  J.~E.~Gerken, A.~Kleinschmidt and O.~Schlotterer,
  ``Heterotic-string amplitudes at one loop: modular graph forms and relations to open strings,''
  arXiv:1811.02548 [hep-th].
  

\bibitem{Broedel:2014vla} 
  J.~Broedel, C.~R.~Mafra, N.~Matthes and O.~Schlotterer,
  ``Elliptic multiple zeta values and one-loop superstring amplitudes,''
  JHEP {\bf 1507}, 112 (2015)
  [arXiv:1412.5535 [hep-th]].
  
  
\bibitem{Green:2013bza} 
  M.~B.~Green, C.~R.~Mafra and O.~Schlotterer,
  ``Multiparticle one-loop amplitudes and S-duality in closed superstring theory,''
  JHEP {\bf 1310}, 188 (2013)
  [arXiv:1307.3534 [hep-th]]. 

\bibitem{Lee:2017ujn} 
  S.~Lee and O.~Schlotterer,
  ``Fermionic one-loop amplitudes of the RNS superstring,''
  JHEP {\bf 1803}, 190 (2018)
  [arXiv:1710.07353 [hep-th]].

 
  
  

  
  
\bibitem{DHokerLectureNotes} 
  E.~D'Hoker
  ``Modular Forms in Physics,'' Lecture notes,
  unpublished.


\bibitem{Green:1997di} 
  M.~B.~Green and P.~Vanhove,
  ``D instantons, strings and M theory,''
  Phys.\ Lett.\ B {\bf 408}, 122 (1997)
  [hep-th/9704145].
  
  
\bibitem{Green:1997as} 
  M.~B.~Green, M.~Gutperle and P.~Vanhove,
  ``One loop in eleven-dimensions,''
  Phys.\ Lett.\ B {\bf 409}, 177 (1997)
  [hep-th/9706175].
 
\bibitem{Green:1999pu} 
  M.~B.~Green, H.~h.~Kwon and P.~Vanhove,
  ``Two loops in eleven-dimensions,''
  Phys.\ Rev.\ D {\bf 61}, 104010 (2000)
  [hep-th/9910055].
   
\bibitem{Green:1999pv} 
  M.~B.~Green and P.~Vanhove,
  ``The Low-energy expansion of the one loop type II superstring amplitude,''
  Phys.\ Rev.\ D {\bf 61}, 104011 (2000)
  [hep-th/9910056].

  
\bibitem{DHoker:2005vch} 
  E.~D'Hoker and D.~H.~Phong,
  ``Two-loop superstrings VI: Non-renormalization theorems and the 4-point function,''
  Nucl.\ Phys.\ B {\bf 715}, 3 (2005)
  [hep-th/0501197].
   
\bibitem{DHoker:2005jhf} 
  E.~D'Hoker, M.~Gutperle and D.~H.~Phong,
  ``Two-loop superstrings and S-duality,''
  Nucl.\ Phys.\ B {\bf 722}, 81 (2005)
  [hep-th/0503180].
  

  
\bibitem{Green:1997tv} 
  M.~B.~Green and M.~Gutperle,
  ``Effects of D instantons,''
  Nucl.\ Phys.\ B {\bf 498}, 195 (1997)
  [hep-th/9701093].
 
\bibitem{DHoker:2014oxd} 
  E.~D'Hoker, M.~B.~Green, B.~Pioline and R.~Russo,
  ``Matching the $D^{6}R^{4}$ interaction at two-loops,''
  JHEP {\bf 1501}, 031 (2015)
  [arXiv:1405.6226 [hep-th]].
  
 
    
\bibitem{DHoker:2015gmr} 
  E.~D'Hoker, M.~B.~Green and P.~Vanhove,
  ``On the modular structure of the genus-one Type II superstring low energy expansion,''
  JHEP {\bf 1508}, 041 (2015)
  [arXiv:1502.06698 [hep-th]].
  
  
\bibitem{Green:2008uj} 
  M.~B.~Green, J.~G.~Russo and P.~Vanhove,
  ``Low energy expansion of the four-particle genus-one amplitude in type II superstring theory,''
  JHEP {\bf 0802}, 020 (2008)
  [arXiv:0801.0322 [hep-th]].
  
  
\bibitem{DHoker:2015wxz} 
  E.~D'Hoker, M.~B.~Green, {\"O}.~G{\"u}rdogan and P.~Vanhove,
  ``Modular Graph Functions,''
  Commun.\ Num.\ Theor.\ Phys.\  {\bf 11}, 165 (2017)
  [arXiv:1512.06779 [hep-th]].
   


\bibitem{DHoker:2016quv} 
  E.~D'Hoker and J.~Kaidi,
  ``Hierarchy of Modular Graph Identities,''
  JHEP {\bf 1611}, 051 (2016)
  [arXiv:1608.04393 [hep-th]].

\bibitem{Basu:2016xrt} 
  A.~Basu,
  ``Poisson equation for the three loop ladder diagram in string theory at genus one,''
  Int.\ J.\ Mod.\ Phys.\ A {\bf 31}, no. 32, 1650169 (2016)
  [arXiv:1606.02203 [hep-th]].
     
\bibitem{Basu:2016kli} 
  A.~Basu,
  ``Proving relations between modular graph functions,''
  Class.\ Quant.\ Grav.\  {\bf 33}, no. 23, 235011 (2016)
  [arXiv:1606.07084 [hep-th]].
    
\bibitem{Basu:2016mmk} 
  A.~Basu,
  ``Simplifying the one loop five graviton amplitude in type IIB string theory,''
  Int.\ J.\ Mod.\ Phys.\ A {\bf 32}, no. 14, 1750074 (2017)
  [arXiv:1608.02056 [hep-th]].
    
\bibitem{Basu:2017nhs} 
  A.~Basu,
  ``Low momentum expansion of one loop amplitudes in heterotic string theory,''
  JHEP {\bf 1711}, 139 (2017)
  [arXiv:1708.08409 [hep-th]].
    
\bibitem{Kleinschmidt:2017ege} 
  A.~Kleinschmidt and V.~Verschinin,
  ``Tetrahedral modular graph functions,''
  JHEP {\bf 1709}, 155 (2017)
  [arXiv:1706.01889 [hep-th]].
    
\bibitem{Basu:2017zvt} 
  A.~Basu,
  ``A simplifying feature of the heterotic one loop four graviton amplitude,''
  Phys.\ Lett.\ B {\bf 776}, 182 (2018)
  [arXiv:1710.01993 [hep-th]].
  
\bibitem{Broedel:2018izr} 
  J.~Broedel, O.~Schlotterer and F.~Zerbini,
  ``From elliptic multiple zeta values to modular graph functions: open and closed strings at one loop,''
  arXiv:1803.00527 [hep-th].
  

  
\bibitem{Brown:2017qwo} 
  F.~Brown,
  ``A class of non-holomorphic modular forms I,''
  arXiv:1707.01230 [math.NT].
  
\bibitem{Brown:2017p2} 
  F.~Brown,
  ``A class of non-holomorphic modular forms II: Equivariant iterated Eisenstein integrals,''
  arXiv:1708.03354 [math.NT].
  
    
\bibitem{DHoker:2017zhq} 
  E.~D'Hoker and W.~Duke,
  ``Fourier series of modular graph functions,''
  arXiv:1708.07998 [math.NT].
    
\bibitem{Zerbini:2018hgs} 
  F.~Zerbini,
  ``Modular and holomorphic graph function from superstring amplitudes,''
  arXiv:1807.04506 [math-ph].
  

\bibitem{DHoker:2019txf} 
  E.~D'Hoker and J.~Kaidi,
  ``Modular graph functions and odd cuspidal functions - Fourier and Poincar\'e series,''
  arXiv:1902.04180 [hep-th].
  
\bibitem{DHoker:2018mys} 
  E.~D'Hoker, M.~B.~Green and B.~Pioline,
  ``Asymptotics of the $D^8 {\cal R}^4$ genus-two string invariant,''
  arXiv:1806.02691 [hep-th].
  
\bibitem{DHoker:2017pvk} 
  E.~D'Hoker, M.~B.~Green and B.~Pioline,
  ``Higher genus modular graph functions, string invariants, and their exact asymptotics,''
  arXiv:1712.06135 [hep-th].
  
\bibitem{Basu:2018eep} 
  A.~Basu,
  ``Supergravity limit of genus two modular graph functions in the worldline formalism,''
  arXiv:1803.08329 [hep-th].

\end{thebibliography}
\end{document}